\documentclass[11pt]{article}
\usepackage[dvips]{epsfig}
\usepackage{epsf}
\begin{document}
\begin{titlepage}

\hfill{\Large{hep-ph/9805216}}

\hfill{\Large{SLAC-PUB 7652}}

\hfill{\Large{DTP-98-18}}

\begin{center}
{\Large \bf Massive Fermionic Corrections to the Heavy Quark Potential}

{\Large \bf Through Two Loops}

\vskip .8in 

{\large Michael Melles \footnote{Present address: University of Durham, UK;
supported by Deutsche Forschungsgemeinschaft, Reference \# Me 1543/1-1
and the European Union TMR-fund.}}

\vskip .4in 

Stanford Linear Accelerator Center

Stanford University, Stanford, CA

\end{center}

\vskip .8in 

\abstract{A physically defined effective charge can incorporate quark masses
analytically at the flavor thresholds. Therefore, no matching conditions
are required for the evolution of the strong coupling constant through
these thresholds. In this paper,
we calculate the massive fermionic corrections to the heavy quark
potential through two loops. The calculation uses a mixed approach of
analytical, computer-algebraic and numerical tools including Monte
Carlo integration of finite
terms. Strong consistency checks are performed by
ensuring the proper cancellation of all non-local divergences by the 
appropriate counterterms and by comparing with the massless limit. 
The size of the effect for the (gauge invariant) fermionic part of $\alpha_V ( {\bf q^2},m^2 ) $
relative to the massless case at the charm and
bottom flavor thresholds is found to be of order $33 \%$.}

\end{titlepage}

\section{Introduction} \label{sec:intro}

In analogy to Quantum Electrodynamics, the heavy quark potential has been of
interest in QCD from very early on \cite{Susskind77,App77,App78,Fischler77,Feinberg78,Billoire80} as a model for the physical definition
of the strong coupling constant \cite{Brodsky83}. Since it represents a potentially
measurable quantity and gives naturally rise to a physical effective
charge $\alpha_V$ \cite{Brodsky83}, it is very interesting to study the QCD flavor thresholds
in such a system \cite{Melles98} as the fermionic corrections are separately gauge
invariant.

In the $MS$ and the
$\overline{MS}$ schemes, the running of the coupling constant, by construction,
does not know about masses of quarks and since the couplings are 
non-physical, the Appelquist-Carazzone \cite{App75} decoupling theorem is not 
applicable.
One has to turn to effective
descriptions which match theories with $m$ massless flavors onto a theory
with $m-1$ massless and one massive flavor at the ``heavy" quark threshold
\cite{Marciano84,Weinberg80,Hall81}. In this
way, the dependence on the
dimensional regularization mass parameter $\mu$ is reduced to next to leading order effects by
giving up the analyticity of the coupling at the flavor threshold \cite{Wetzel82,Bernreuther82,Bernreuther83,Bernreuther83a,Chetyrkin97,Rodrigo93}. 

While this procedure of matching conditions and effective descriptions is 
certainly workable, from a theoretical standpoint it would be advantageous
to have a physical coupling constant definition which is analytic at thresholds.
In addition, as a physical observable, the total derivative with respect to the
renormalization scale $\mu$ vanishes.
Such a system is given by identifying the ground state energy of the vacuum 
expectation value of the Wilson loop as
the potential $V$ between a static quark-antiquark pair in a
color singlet state \cite{Susskind77,Fischler77,Peter97}:

\begin{equation}
V( r, m^2) = - \lim_{t \rightarrow \infty} \frac{1}{it} log \langle 0| Tr \; \{
P \; exp \left( \oint dx_\mu A^\mu_a T^a \right) \} |0 \rangle \label{eq:Vdef}
\end{equation}

where $r$ denotes the relative distance between the heavy quarks, $m$ the mass
of ``light'' quarks contributing through loop effects and $T^a$ the generators
of the gauge group. It is then convenient to define the effective charge
$\alpha_V( {\bf q}^2,m^2)$ as

\begin{equation}
V({\bf q}^2,m^2) \equiv - \frac{4 \pi C_F \alpha_V ( {\bf q}^2,m^2)}{{\bf q}^2}
\label{eq:aVdef}
\end{equation}

in momentum space. The factor $C_F$ is the value of the Casimir operator $T^aT^a$
in the fundamental representation of the external sources and factors out to all
orders in perturbation theory. As one is free to choose the representation 
of the external particles, we obtain the static gluino potential by adopting the
adjoint representation. 

The massless case was recently calculated in Ref. \cite{Peter96} and in this 
paper, we will give all the two loop fermionic contributions to $\alpha_V
({\bf q}^2,m^2)$ for all perturbative values of the momentum transfer
${\bf q}^2 \equiv q_0^2-q^2=-q^2 >0$
and for arbitrary values of the fermion mass $m$.
In this context we are only interested in the two loop
contributions to the potential in the effective Schroedinger equation for the
heavy particles. This implies, for instance, that not always the whole diagram
contributes to the potential as certain parts can already be reproduced 
by the exponentiation of lower order diagrams. The necessity for this subtlety 
has its origin in the exponential present in Eq. \ref{eq:Vdef}.
For a detailed discussion, see
Ref. \cite{Fischler77}.

It is also important to note that the results of massive two loop integrals 
presented in this work are also relevant for the related problem of 
quark threshold production. For this application, though, it would be 
necessary to treat also the occurring imaginary parts of the integrals
numerically as pole terms will contribute for timelike momentum transfers
at the production threshold $q^2=4m^2$. A promising approach for this treatment
might be the recently suggested Taylor expansion of integrands around threshold
\cite{Beneke97} by determining large and small scales in the problem. 
The heavy quark approximation eliminates the possibility of timelike momentum
transfers in this work so that we do not need to worry about pole terms
numerically. Nevertheless, we also list the contributions needed in this case
for all integrals.

The paper is outlined as follows:

In section \ref{sec:nac} we list all the occurring two loop contributions
explicitly in the Feynman gauge and with the usage of heavy quark effective
Feynman rules for the external sources.
In section \ref{sec:res} the unrenormalized results for the two loop corrections
are given in terms of two loop scalar integrals, for which explicit expressions
are listed in appendix \ref{sec:tli}.
Section \ref{sec:ren} contains all the required counterterms in the 
$MS$-renormalization scheme and it is shown
that all non-local divergences cancel. The renormalization constants obtained
are given explicitly and checked with the known results.
Section \ref{sec:numres} contains numerical results which demonstrate that
the massless limit is obtained correctly and display the effect of including
the mass terms for the charm and bottom flavor thresholds. 
In section \ref{sec:con} we make concluding remarks and indicate future
lines of work with the presented results. Appendix \ref{sec:decomp}, finally, lists all the
reductions from tensor to scalar integrals needed for the results displayed in 
section \ref{sec:res}.
  
\section{The Two Loop Corrections} \label{sec:nac}

\begin{center}
\begin{figure}
\centering
 \epsfig{file=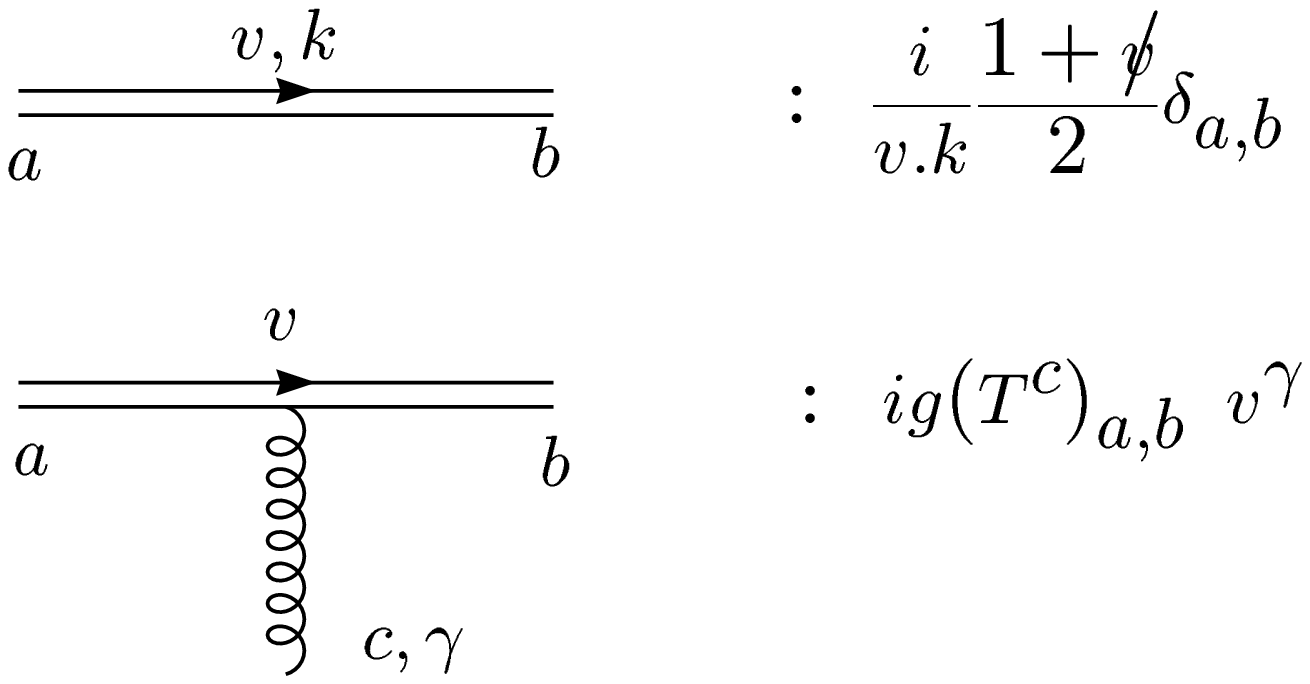,height=4cm}
\caption{The Feynman rules for heavy quark effective theory used in this
work for the source propagator and the source gluon vertex. For anti sources one
has to make the replacement $v \longrightarrow - v $. The $i$-$\varepsilon$ 
prescription is the same as for the usual fermion propagator.}
 \label{fig:hqeFr}
\end{figure}
\end{center}

\begin{center}
\begin{figure}
\centering
 \epsfig{file=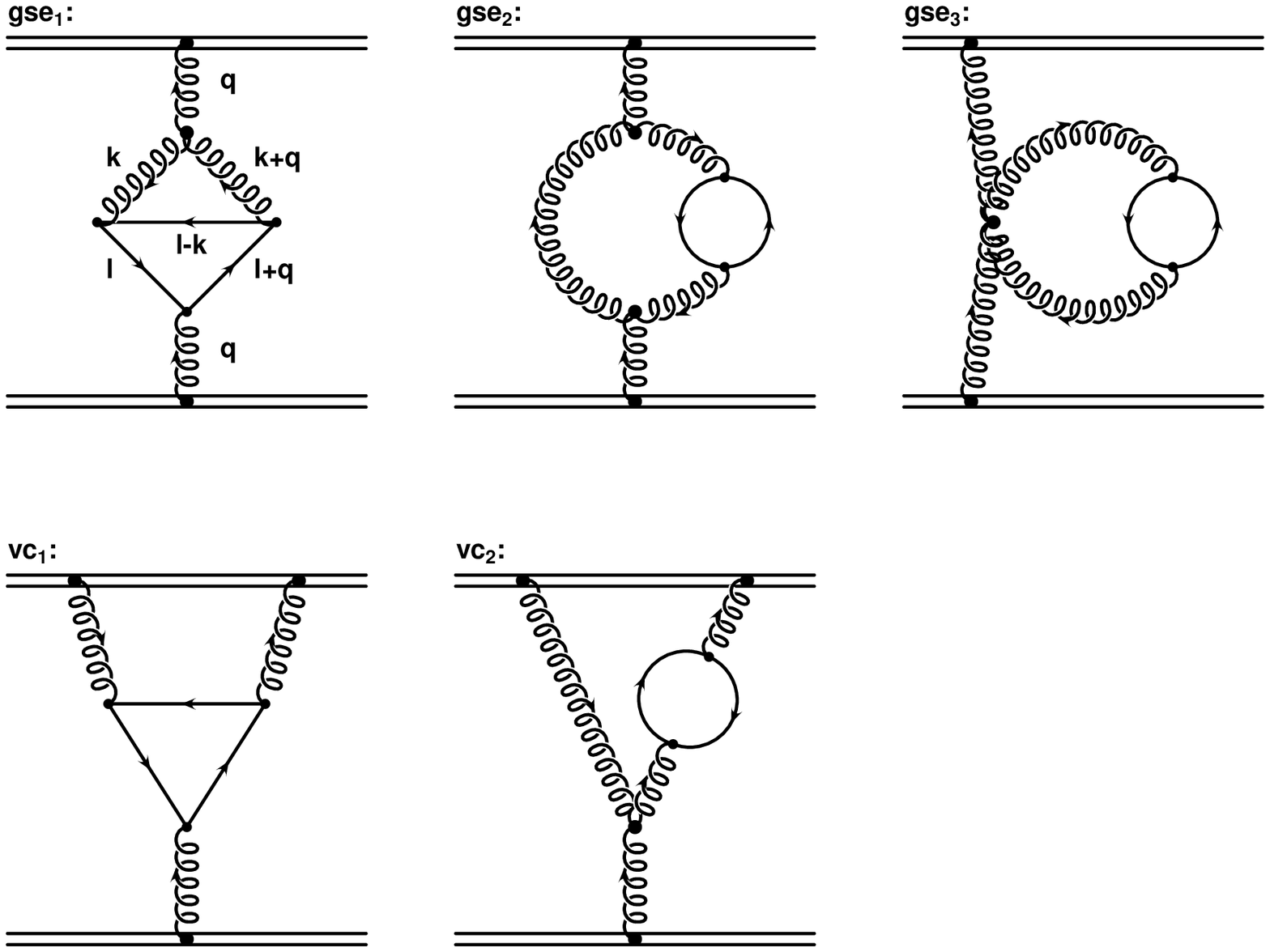,width=14cm}
 \vspace{1.3cm} \\
 \epsfig{file=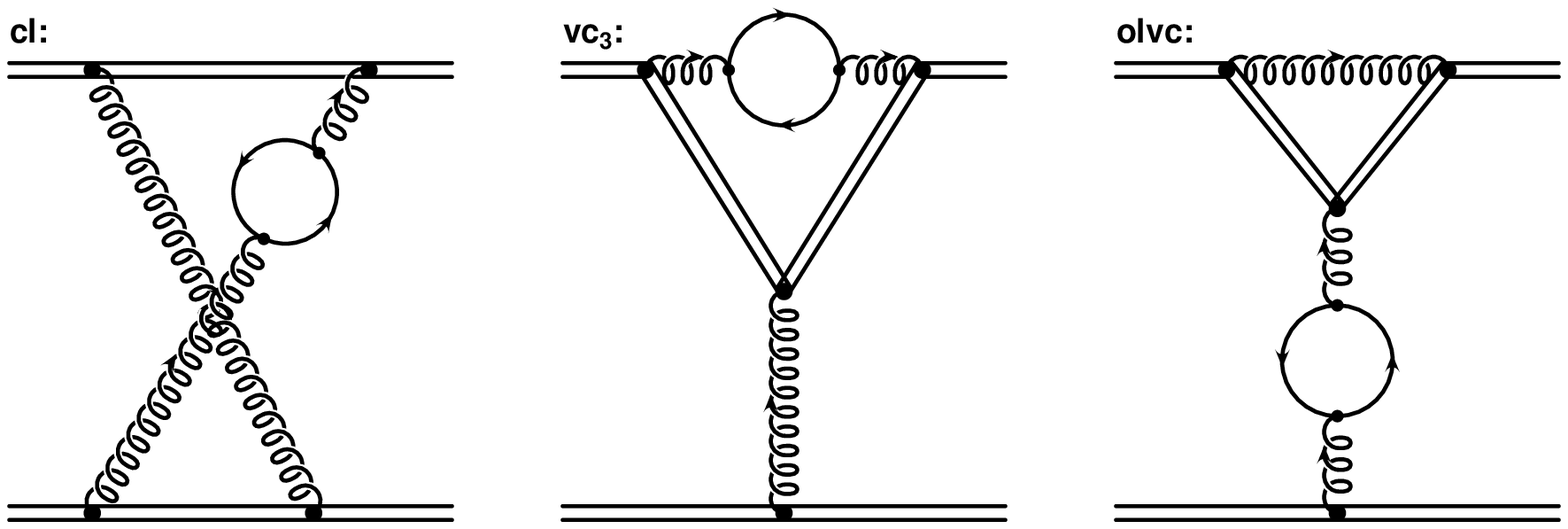,width=14cm}
 \caption{The non-Abelian Feynman diagrams contributing to the massive
fermionic corrections to the heavy quark potential at the two loop level. 
The first two rows contain diagrams with a typical non-Abelian topology.
Double lines denote the heavy quarks, single lines the ``light'' quarks. Color
and Lorentz indices are suppressed in the first graph. The notation for
the remaining digrams is analogous.
The last line includes the infra-red divergent ``Abelian'' Feynman diagrams.
While the topology of these three diagrams is the same as in QED, they
contribute to the potential only in the non-Abelian theory due to
color factors ~$C_F C_A$. In addition, although each diagram is infra-red 
divergent, their sum is infra-red finite.}
 \label{fig:hqpfig1}
\end{figure}
\end{center}

\begin{center}
\begin{figure}[t]
\centering
 \epsfig{file=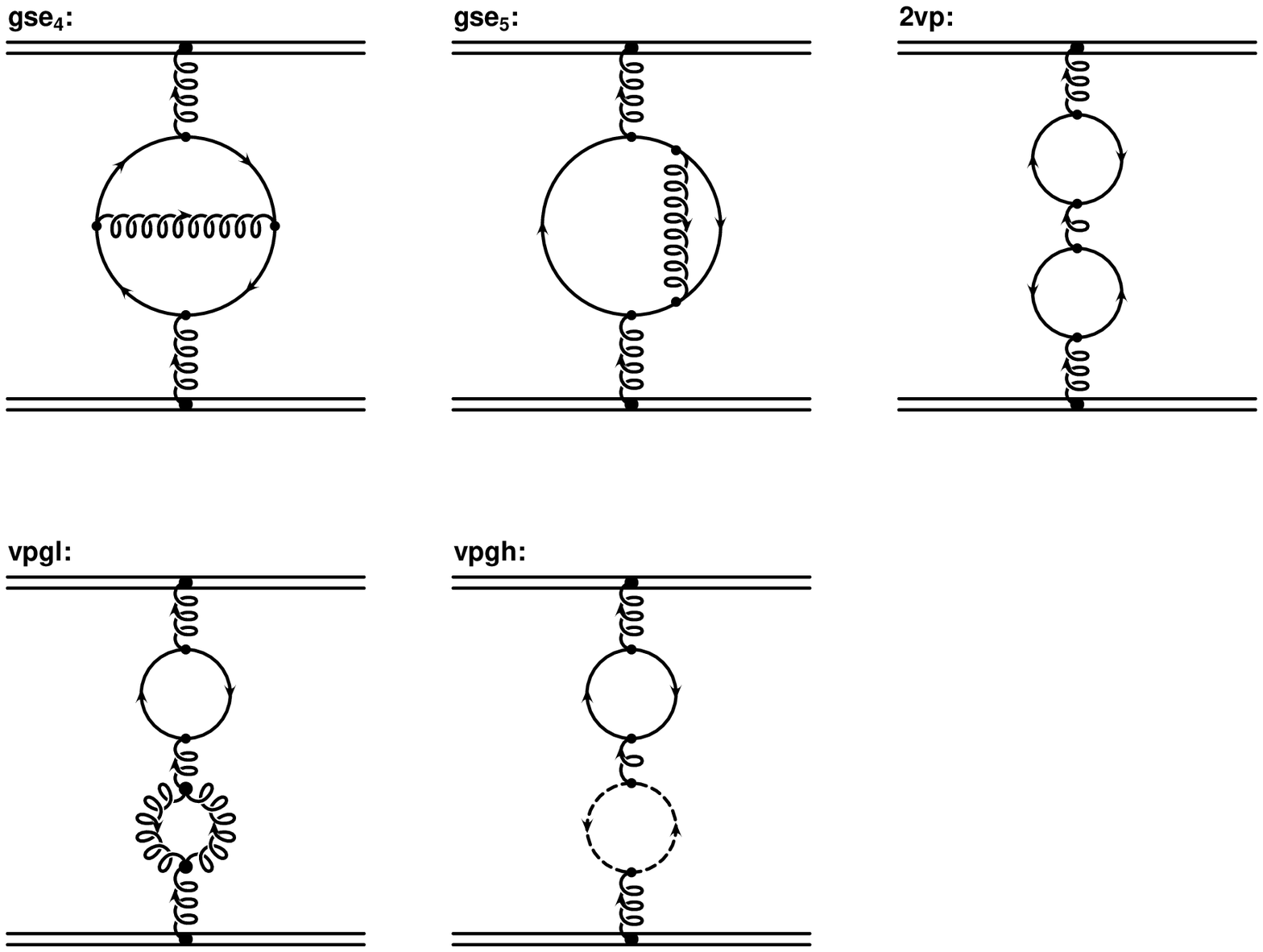,height=10cm}
 \caption{The infra-red finite Feynman diagrams with an Abelian topology 
(upper line) contributing to the massive
fermionic corrections to the heavy quark potential at the two loop level
plus diagrams consisting of one loop insertions with non-Abelian terms
(lower line).} 
 \label{fig:hqpfig1a}
\end{figure}
\end{center}

In this section we present the non-Abelian contributions to the heavy quark
potential that constitute the new results of this work. They are depicted in Fig.
\ref{fig:hqpfig1}.
The $QED$ like 
diagrams, which need to be modified by their respective color factors, 
have been known for a long time \cite{Sabry55} and can also be found in Refs. 
\cite{Chetyrkin97,Hoang94,Kniehl90,Barbieri73} for instance. They are given here as well because we would like
to be able to separate non-Abelian and Abelian contributions to the potential.
It has been observed before \cite{Chetyrkin97} that their respective threshold
behavior can be quite different. These diagrams, together with effectively
``one loop'' diagrams are given in Fig. \ref{fig:hqpfig1a}. 
The weighted sum of all
the graphs shown, modulo terms already generated by the exponentiation of
the lower order Born and the one loop vacuum polarization diagram, 
give the complete gauge invariant fermionic corrections to the heavy quark potential
at two loops in the Feynman gauge. The choice of this gauge simplifies the calculation
because the decomposition into scalar two loop integrals is easier and it also 
reduces the three gluon vertex correction graph to zero in the 
heavy quark effective theory.
Below we list all contributions
at the two loop level. The abbreviations stand for
$gse \equiv gluon \; self \; energy$, $vc \equiv vertex \; correction$,
$cl \equiv crossed \; ladder$ and $olvc \equiv one \; loop \; vertex \;
correction$.
In the heavy quark limit we use the source gluon vertex and 
source propagator Feynman rules
of heavy quark effective theory \cite{Neubert94,Buchalla94} which are given in Fig. \ref{fig:hqeFr}. 

With these, and taking $v_\mu \equiv (1,0,0,0)$ and $q_0=0$ for the purely 
spacelike momentum transfer $q$,
the two loop diagrams of Figs. \ref{fig:hqpfig1} and 
\ref{fig:hqpfig1a} read in the Feynman gauge (summed
over the external color degrees of freedom and including a symmetry factor of 
$\frac{1}{2}$ for the first three amplitudes):

\begin{eqnarray}
{\cal M}_{gse_1} &\equiv& \frac{-ig^6\mu^{2\epsilon}}{q^4} \frac{C_F C_A T_F}{4} \delta^{\alpha,0} 
\delta^{\beta,0} \int \frac{d^nk}{(2 \pi)^n} \int 
\frac{d^nl}{(2 \pi)^n} \biggl[ \frac{Tr \left\{ \gamma^\delta \left( \rlap / l - \rlap / k
+m \right) \gamma^\gamma \left( \rlap / l + \rlap / q + m \right) \gamma_\alpha
\left( \rlap / l + m \right) \right\}}{((l+q)^2-m^2)(l^2-m^2)((l-k)^2-m^2)(k+q)^2
k^2} \nonumber \\
&& \!\!\!\!\!\!\!\!\!\!\!\! \times \left( (q-k)_\gamma g_{\delta,\beta} + (-k-2q)_\delta g_{\gamma,
\beta} + (2k + q )_\beta g_{\delta,\gamma} \right) \biggr] \label{eq:gse1def} \\
{\cal M}_{gse_2} &\equiv& \frac{-ig^6\mu^{2\epsilon}}{q^4} \frac{C_F C_A T_F}{2} \delta^{\alpha,0} 
\delta^{\beta,0} \int \frac{d^nk}{(2 \pi)^n} \int 
\frac{d^nl}{(2 \pi)^n} \biggl[ \frac{Tr \left\{
\gamma^\gamma \left( \rlap / l - \rlap / k + m \right) \gamma^\delta
\left( \rlap / l + m \right) \right\}}{(l^2-m^2)((l-k)^2-m^2)(k+q)^2
k^4} \nonumber \\
&& \!\!\!\!\!\!\!\!\!\!\!\! \times \left( (-2q-k)_\gamma g_{\sigma,\alpha} + (-k+q)_\sigma g_{\gamma,
\alpha} + (2k + q )_\alpha g_{\sigma,\gamma} \right) \left( (q-k)_\sigma 
g_{\delta,\beta} + (-2q-k)_\delta g_{\sigma,\beta} + (2k+q)_\beta 
g_{\delta,\sigma}
\right) \biggr] \label{eq:gse2def} \\
{\cal M}_{gse_3}  &\equiv& \frac{-ig^6\mu^{2\epsilon}}{q^4} \frac{C_F C_A T_F}{2} \delta^{\alpha,0} 
\delta^{\beta,0}  \int \frac{d^nk}{(2 \pi)^n} \int 
\frac{d^nl}{(2 \pi)^n} \biggl[ \frac{Tr \left\{
\gamma^\gamma \left( \rlap / l - \rlap / k + m \right) \gamma^\delta
\left( \rlap / l + m \right) \right\}}{(l^2-m^2)((l-k)^2-m^2)k^4} \nonumber \\
&& \!\!\!\!\!\!\!\!\!\!\!\! \times
\left( g_{\gamma,\beta} g_{\alpha,\delta} - 2 g_{\gamma,
\delta} g_{\alpha,\beta} + g_{\gamma,\alpha} g_{\delta,\beta} \right) \biggr]
\label{eq:gse3def} \\
{\cal M}_{gse_4}  &\equiv& \frac{-ig^6\mu^{2\epsilon}}{q^4} \left( C_F^2 - \frac{C_F C_A}{2} \right) T_F \; \delta^{\alpha,0} 
\delta^{\beta,0}  \int \frac{d^nk}{(2 \pi)^n} \int \frac{d^nl}{(2 \pi)^n} \nonumber \\
&& \times \biggl[ \frac{Tr \left\{
\gamma_\alpha \left( \rlap / l + \rlap / q + m \right) \gamma_\gamma
\left( \rlap / k + \rlap / q +  m \right) \gamma_\beta 
\left( \rlap / k +  m \right) \gamma^\gamma \left( \rlap / l +  m \right)
\right\}}{((l+q)^2-m^2)(l^2-m^2)(l-k)^2((k+q)^2-m^2)(k^2-m^2)} 
\biggr] \label{eq:gse4def} \\
{\cal M}_{gse_5}  &\equiv& \frac{-ig^6\mu^{2\epsilon}}{q^4} C_F^2 T_F \; \delta^{\alpha,0} 
\delta^{\beta,0}  \int \frac{d^nk}{(2 \pi)^n} \int \frac{d^nl}{(2 \pi)^n} 
\biggl[ \frac{Tr \left\{
\gamma_\alpha \left( \rlap / k + \rlap / q + m \right) \gamma_\beta
\left( \rlap / k +  m \right) \gamma_\gamma
\left( \rlap / l +  m \right) \gamma^\gamma \left( \rlap / k +  m \right)
\right\}}{(l^2-m^2)(l-k)^2((k+q)^2-m^2)(k^2-m^2)^2} 
\biggr] \label{eq:gse5def} \\
{\cal M}_{vc_1} &\equiv& \frac{ig^6\mu^{2\epsilon}}{q^2} \frac{C_F C_A T_F}{2} \delta^{\alpha,0} 
\delta^{\beta,0} \delta^{\gamma,0} \int \frac{d^nk}{(2 \pi)^n} \int 
\frac{d^nl}{(2 \pi)^n} \frac{Tr \left\{ \gamma_\gamma \left( \rlap / l - \rlap / k
+m \right) \gamma_\alpha \left( \rlap / l + \rlap / q + m \right) \gamma_\beta
\left( \rlap / l + m \right) \right\}}{((l+q)^2-m^2)(l^2-m^2)((l-k)^2-m^2)
(k+q)^2k^2 k_0} \label{eq:vc1def} \\
{\cal M}_{vc_2} &\equiv& \frac{ig^6\mu^{2\epsilon}}{q^2} \frac{C_F C_A T_F}{2} \delta^{\alpha,0} 
\delta^{\beta,0} \delta^{\gamma,0} \int \frac{d^nk}{(2 \pi)^n} \int 
\frac{d^nl}{(2 \pi)^n} \biggl[ \frac{Tr \left\{
\gamma_\alpha \left( \rlap / l - \rlap / k + m \right) \gamma_\nu
\left( \rlap / l + m \right) \right\}}{(l^2-m^2)((l-k)^2-m^2)(k+q)^2k^4k_0}
\nonumber \\
&& \!\!\!\!\!\!\!\!\!\!\!\! \times 
\left( (q-k)_\beta g_{\nu,\gamma}+(-k-2q)_\nu g_{\beta,\gamma}+ (2k + q)_\gamma
g_{\nu,\beta} \right) \biggr] \label{eq:vc2def} \\
{\cal M}_{vc_3} &\equiv& \frac{ig^6\mu^{2\epsilon}}{q^2} \frac{C_F C_A T_F}{2} \delta^{\alpha,0} 
\delta^{\beta,0} \int \frac{d^nk}{(2 \pi)^n} \int 
\frac{d^nl}{(2 \pi)^n} \frac{Tr \left\{
\gamma_\alpha \left( \rlap / l - \rlap / k + m \right) \gamma_\beta
\left( \rlap / l + m \right) \right\}}{(l^2-m^2)((l-k)^2-m^2)k^4(k_0+i\varepsilon)^2}
\label{eq:vc3def} \\
{\cal M}_{cl} &\equiv& -ig^6\mu^{2\epsilon} \frac{C_F C_A T_F}{2} \delta^{\alpha,0} 
\delta^{\beta,0} \int \frac{d^nk}{(2 \pi)^n} \int 
\frac{d^nl}{(2 \pi)^n} \frac{Tr \left\{
\gamma_\alpha \left( \rlap / l - \rlap / k + m \right) \gamma_\beta
\left( \rlap / l + m \right) \right\}}{(l^2-m^2)((l-k)^2-m^2)k^4(k+q)^2(k_0+i\varepsilon)^2}
\label{eq:cldef} \\
{\cal M}_{olvc} &\equiv& \frac{ig^6\mu^{2\epsilon}}{q^4} \frac{C_F C_A T_F}{2} \delta^{\alpha,0} 
\delta^{\beta,0} \int 
\frac{d^nl}{(2 \pi)^n} \frac{Tr \left\{
\gamma_\alpha \left( \rlap / l - \rlap / q + m \right) \gamma_\beta
\left( \rlap / l + m \right) \right\}}{(l^2-m^2)((l-q)^2-m^2)}\int \frac{d^nk}{(2 \pi)^n}
\frac{1}{k^2(k_0+i\varepsilon)^2}
\label{eq:olvcdef}  
\end{eqnarray}

It should be noted that in our case there is no need for an $i$-$\varepsilon$ 
prescription in the denominators of Eqs. \ref{eq:gse1def} through \ref{eq:vc2def}
as the spacelike nature of the physical momentum transfer only leads to purely real integrals and no unambiguous pole terms occur in the denominators of 
those diagrams. 
This feature also simplifies the Monte Carlo integration of the finite parts 
of the contributing graphs. The three graphs \ref{eq:vc3def}, \ref{eq:cldef}
and \ref{eq:olvcdef} display infra-red 
divergences which cancel in the sum. 
The one loop vertex correction graph ${\cal M}_{olvc}$ vanishes in dimensional
regularization, however, is needed to ensure the proper cancellation of infra-red
divergences.

The color factors given are not always 
the full color factors. Only those contributing to the potential are listed.
The Casimir invariants \cite{PeskSchr} for a general $SU(N)$ group are defined by

\begin{equation}
C_A \equiv N \;\;\;\;\;,\;\;\;\;\; C_F \equiv \frac{N^2-1}{2N}
\end{equation}

Furthermore, $Tr\{T^aT^b\} \equiv T_f \delta^{a,b} = \frac{1}{2} \delta^{a,b}$.
The color factor for ${\cal M}_{vc_1}$ includes the sum of the graph shown in Fig. 
\ref{fig:hqpfig1} plus the term stemming from the fermion momenta reversed
contribution. Only the sum is proportional to $C_A$, the other terms vanish
according to Furry's theorem, as is the case in QED. 
For QCD, the crossed ladder diagrams do contribute as they contain a color 
factor proportional to $C_F^2-\frac{C_FC_A}{2}$, whereas the straight ladder
graph has a color factor proportional to $C_F^2$ only. This will be expounded
on in section \ref{sec:IR}.
In QED, the sum of all vertex, ladder and crossed ladder Feynman diagrams are
equivalent to the iteration of the potential in the Schroedinger theory.
$\alpha_{V_{QED}}$
and the effective coupling \cite{DeRafael73} differ, therefore, only at three loops 
due to
light by light scattering contributions. 

\section{Unrenormalized Results} \label{sec:res}

The two loop integrals needed for the expressions of Eqs. \ref{eq:gse1def}
through \ref{eq:vc2def} are treated in separate ways in
this work depending on whether or not they contain two or more internal
fermion lines. In the former case we integrate the fermion loop first as will
be explained below. For the vertex correction contribution ${\cal M}_{vc_1}$
we integrate the fermion loop analytically as well with all the Lorentz indices
projected to zero and then proceed with additional Feynman parameters for the
remaining loop integration. 

The two point functions ${\cal M}_{gse_1}$, ${\cal M}_{gse_4}$ and
${\cal M}_{gse_5}$ are treated in a
completely different manner as the above techniques would now be too cumbersome. We project the complicated tensor structure onto scalar quantities as described
below and then proceed with an algebraic reduction into scalar two loop
integrals. This reduction is programmed in FORM \cite{verma} and details are
presented in appendix \ref{sec:decomp}. The resulting scalar integrals are
then evaluated by employing standard Feynman parameter techniques and explicit
results are listed in appendix \ref{sec:tli}. Overall results for the various
amplitudes are obtained by expanding the n-dimensional results around $\epsilon
=0$ with MAPLE. It is important to notice, given the complexity of the 
calculation, that the translation into FORTRAN code was also performed by MAPLE,
thus dramatically reducing the chance of accidental mistakes.
The evaluation of finite parts
is done with the Monte Carlo integrator VEGAS \cite{Lepage}.

For the two point functions we use the following decomposition into transverse
($t$) and longitudinal ($l$) components:

\begin{equation}
\Pi_{\alpha,\beta} \left( q \right) \equiv \left(  g_{\alpha,\beta} -
\frac{q_\alpha q_\beta}{q^2} \right) \Pi_t \left( q^2 \right) + \frac{q_\alpha q_\beta}
{q^2} \Pi_l \left( q^2 \right) \label{eq:Pidecomp}
\end{equation}

from which it follows that in $n=4-\epsilon$ dimensions

\begin{eqnarray}
\Pi_t \left( q^2 \right) &=& \frac{1}{n-1} \left(  g^{\alpha,\beta} - 
\frac{q^\alpha q^\beta}{q^2}
\right) \Pi_{\alpha,\beta} \left( q \right) \label{eq:pitres} \\
\Pi_l \left( q^2 \right) &=&  \frac{q^\alpha q^\beta}{q^2} \Pi_{\alpha,\beta} 
\left( q \right) \label{eq:pilres}
\end{eqnarray}

With this notation  and the heavy quark effective Feynman rules depicted in
Fig. \ref{fig:hqeFr} we arrive at

\begin{equation}
{\cal M}_{gse_i} \equiv \frac{g^2C_F}{q^4} \delta^{\alpha,0} \delta^{\beta,0} \Pi^i_{\alpha,
\beta} \left( q \right) \;\;\;\;\;, \; i = \{1...5\} \label{eq:gseidef}
\end{equation}

The result of the decomposition for the transverse component of the gluon
self energy graph ${\cal M}_{gse_1}$, using the relations given in appendix \ref{sec:decomp}, reads

\begin{eqnarray}
\Pi^1_t &=& \frac{i g^4 C_A T_F}{4(n-1)} \left[ \left( n \frac{8}{3} - \frac{20}
{3} \right)  \left( A_2 B_{12'} 
-m^2 T_{12'35} \right) + \left( 4n-10 \right) T_{235} + \left( 8 -4n \right)
A_2 B_{12} + \left( n \frac{8}{3} - \frac{14}{3} \right) T_{135} \right. \nonumber \\
&& + 16 \left(q^2-q^2 \frac{n}{2} - m^2 \right) T_{1235} + q^2 \left( 4 n -6
\right) T_{2345} + q^2 \left( 2 n -4 \right) B_{12} B_{45} -8q^2 m^2 T_{12345}
+ 8 A_2 B_{45} \nonumber \\ && +
\frac{m^2}{q^2} \left\{ \left( n \frac{8}{3} - 
\frac{20}{3} \right)  \left( m^2 T_{12'35} + T_{2'35} - A_2 B_{12'} + 
\frac{1}{m^2} A_2^2 \right) +16 T_{235} - \left( n \frac{8}{3}  
\right. \right. \nonumber \\ && + \left. \left. \frac{28}{3} 
\right) T_{135} \right\}
+ \frac{1}{n-1} \left\{ -n \frac{2}{3} m^2 T_{12'35} +n T_{235}
-n \frac{7}{3} T_{135} -4 n A_2 B_{45} +n \frac{2}{3} A_2 B_{12'} -  
8 m^2 T_{2345} \right. \nonumber \\
&& - 4 m^2 T_{2355} + 4 T_{135} + q^2 \left( 4 m^2 T_{23455}
+ (4-2n) A_2 C_{455}- n T_{2345} \right) + \frac{1}{q^2} \left( 4 m^2 
\left( T_{135}-T_{235} \right) \right. \nonumber \\
&& \left. \left. \left. + n \frac{2}{3} \left( m^4 T_{12'35} - m^2 T_{135}+ m^2 T_{2'35}-m^2 A_2
B_{12'} + A_2^2 \right) \right) \right\} \right]  \label{eq:pi1t}
\end{eqnarray}

It is also useful to examine the $m \longrightarrow 0$ limit of the above 
expression as this case was calculated in Ref. \cite{Peter97} and can serve
a valuable test for the above expression. By inspecting the occurring integrals
we find the massless limit to correspond to

\begin{eqnarray}
\Pi^1_{t_{m \longrightarrow 0}} &=& \frac{i g^4 C_A T_F}{4(n-1)} \left[ 
\left( n \frac{8}{3}- \frac{n}{n-1} \frac{7}{3}  - \frac{14}{3} +\frac{4}{n-1}
\right) T_{135} 
+16 \left(q^2-q^2 \frac{n}{2} \right) T_{1235} 
+ q^2 \left( 2 n -4 \right) B_{12} B_{45} \right. \nonumber \\ 
&& \left. + q^2 \left( 4 n -6 -\frac{n}{n-1}\right) T_{2345} 
 \right] 
\label{eq:pi1tm0}
\end{eqnarray}

These terms are also, as expected, the only ones contributing to the 
gluon wave function renormalization constant. In other words, all divergent
parts of the two  and one loop integrals which vanish in the massless
limit in the expression \ref{eq:pi1t} add up to zero identically. 
This in itself is an important check of the overall expression.
In the heavy quark limit we can neglect the timelike component of the
four momentum transfer $q$, i.e. $ q_0 =0$ as was already mentioned before.
This means that we do not need the longitudinal component of ${\cal M}_{gse_1}$, however,
we list it here for completeness:

\begin{eqnarray}
\Pi^1_l &=& \frac{i g^4 C_A T_F}{4} \left( \left( n \frac{8}{3} - \frac{20}{3} \right) \left( m^2 T_{12'35}
- A_2 B_{12'} + \frac{1}{2} T_{135} \right) + 2 T_{235} - 8 A_2 B_{45} - 2
q^2 T_{2345} \right. \nonumber \\ && +
\frac{m^2}{q^2} \left\{ \left( n \frac{8}{3} - \frac{20}{3} \right) \left(
A_2 B_{12'}- \frac{1}{m^2} A_2^2 - T_{2'35} - m^2 T_{12'35} \right) + \left(
n \frac{8}{3} + \frac{28}{3} \right) T_{135} - 16 T_{235} 
\right\} \nonumber \\ && +
\frac{1}{n-1} \left[ n \frac{2}{3} m^2 T_{12'35} -n T_{235} + n \frac{7}{3}
T_{135} +4 n A_2 B_{45} - n \frac{2}{3} A_2 B_{12'} + 8 m^2 T_{2345} +4 m^2
T_{2355} - 4 T_{135} \right. \nonumber \\ && +
q^2 \left( n T_{2345} + 2 n A_2 C_{455} - 4 m^2 T_{23455} -4 A_2 C_{455}
\right) + \frac{1}{q^2} \left\{ 4m^2 \left( T_{235}-T_{135} \right) \right. \nonumber \\
&& \left. - \left. \left. n \frac{2}{3} \left( m^4 T_{12'35} - m^2 T_{135} + m^2 T_{2'35}
- m^2 A_2 B_{12'} + A_2^2 \right) \right\} \right] \right) \label{eq:pi1l}
\end{eqnarray}

A good check on the consistency of the employed decomposition is given by the
absence of infra-red divergences. None of the two point amplitudes in this work is
infra-red divergent to begin with, however, in intermediate steps of the
calculation those do occur. An example is given above by the two integrals
$T_{2355}$ and $T_{23455}$ for which only the combination $q^2T_{23455}-T_{2355}$
is infra-red finite and this is how they enter into Eqs. \ref{eq:pi1t}
and \ref{eq:pi1l}. The function $A_2C_{455}$ only seems to have an infra-red 
divergence, however, in dimensional regularization it can actually be written
as an ultra-violet divergence. This is done in appendix \ref{sec:tli}.

For the two diagrams that have an Abelian topology, Eqs. \ref{eq:gse4def} and
\ref{eq:gse5def}, we also give explicit results as usually only their sum is
given in the literature \cite{Chetyrkin97,Weiglein94}. Here, however, we need both contributions
separately due to the different color factors. In addition,
Abelian and non-Abelian terms are
separately gauge invariant and might display a different threshold behavior
\cite{Chetyrkin97}. We find:

\begin{eqnarray}
\Pi^4_t &=& \frac{i g^4 (C_F-\frac{C_A}{2}) T_F}{(n-1)} \left[ \left( n \frac{8}{3} - \frac{16}
{3} \right)  \left( A_2 B_{12'} 
-m^2 T_{12'35} \right) 
+\left( q^2 \left( (4 n-8) q^2 + (32 - 8 n) m^2 \right) -32 m^4 \right) T^A_{12345} \right. \nonumber \\
&& + \left( 8n-16 \right) \left( T_{235} -A_2 B_{12} \right) + 32 \left(q^2-q^2 \frac{n}{2} - m^2 \right) T_{1235} 
+ \left(4 n^2 - n \frac{76}{3} + \frac{104}{3} \right) T_{135}
\nonumber \\ && + \left( q^2 \left( 18 n -2 n^2 -28 \right) +16 m^2 \right) B_{12} B_{12}
+ \frac{1}{q^2} \left( 
\frac{8}{3} n  \left( m^4 T_{12'35} - m^2 T_{135}+ m^2 T_{2'35}-m^2 A_2B_{12'}
+ A_2^2 \right) \right. \nonumber \\
&& \left. \left. + 16 m^2 \left( T_{235}-T_{135} \right) - \frac{16}{3} \left( m^4 T_{12'35} - m^2 T_{135}+ m^2 T_{2'35}-m^2 A_2
B_{12'} + A_2^2 \right) \right) \right]  \label{eq:pi4t}
\end{eqnarray}

and for the longitudinal component:

\begin{eqnarray}
\Pi^4_l &=& i g^4 (C_F-\frac{C_A}{2}) T_F \left[ \left( n \frac{8}{3} - \frac{16}
{3} \right)  \left(m^2 T_{12'35} + \frac{1}{2}T_{135} - A_2 B_{12'} \right)
\right. \nonumber \\ && 
+ \frac{1}{q^2} \left( 
\frac{16}{3} \left( m^4 T_{12'35} - m^2 T_{135}+ m^2 T_{2'35}-m^2 A_2B_{12'}
+ A_2^2 \right) \right. \nonumber \\
&& \left. \left. + 16 m^2 \left( T_{135}-T_{235} \right) - \frac{8}{3} n \left( m^4 T_{12'35} - m^2 T_{135}+ m^2 T_{2'35}-m^2 A_2
B_{12'} + A_2^2 \right) \right) \right]  \label{eq:pi4l}
\end{eqnarray}

Similarly, for Eq. \ref{eq:gse5def} we get the following result:

\begin{eqnarray}
\Pi^5_t &=& \frac{i g^4 C_F T_F}{(n-1)} \left[ \left(n^2- n \frac{16}{3} - \frac{20}
{3} \right)  \left( A_2 B_{12'} 
-m^2 T_{12'35} \right) -16 m^2 T_{1235}
+\left( (16 - 8 n) q^2 m^2 -32 m^4 \right) T_{12235} \right. \nonumber \\
&& + \! \left(n^2-4n +4
\right) \!\! \left(q^2 T_{12'35}-T_{2'35}+2 A_2 B_{22}-A_2 B_{22'}+q^2A_2C_{122'} -2q^2 A_2 C_{122} \right) 
 \! - \! \left( n^2 - \frac{14}{3}n + \frac{16}{3} \right) \! T_{135}
\nonumber \\ && + m^2 \left( 8 n -16 \right) \left( T_{2235}-A_2 C_{122} \right)
- \frac{1}{q^2} \left( 
\frac{4}{3} n  \left( m^4 T_{12'35} - m^2 T_{135}+ m^2 T_{2'35}-m^2 A_2B_{12'}
+ A_2^2 \right) \right. \nonumber \\
&& \left. \left. + 8 m^2 \left( T_{235}-T_{135} \right) - \frac{8}{3} \left( m^4 T_{12'35} - m^2 T_{135}+ m^2 T_{2'35}-m^2 A_2
B_{12'} + A_2^2 \right) \right) \right]  \label{eq:pi5t}
\end{eqnarray}

and for the longitudinal component:

\begin{eqnarray}
\Pi^5_l &=& - i g^4 C_F T_F \left[ \left( n \frac{4}{3} - \frac{8}
{3} \right)  \left(m^2 T_{12'35} + \frac{1}{2}T_{135} - A_2 B_{12'} \right)
\right. \nonumber \\ && 
+ \frac{1}{q^2} \left( 
\frac{8}{3} \left( m^4 T_{12'35} - m^2 T_{135}+ m^2 T_{2'35}-m^2 A_2B_{12'}
+ A_2^2 \right) \right. \nonumber \\
&& \left. \left. + 8 m^2 \left( T_{135}-T_{235} \right) - \frac{4}{3} n \left( m^4 T_{12'35} - m^2 T_{135}+ m^2 T_{2'35}-m^2 A_2
B_{12'} + A_2^2 \right) \right) \right]  \label{eq:pi5l}
\end{eqnarray}

It can easily be seen that both parts of the two functions in Eqs. \ref{eq:pi4t}
and \ref{eq:pi5t} multiplying
$\frac{1}{q^2}$ are identical up to a minus sign when \ref{eq:pi5t} is multiplied 
by the multiplicity 
factor $2$. This is required  
by the gauge structure of the gluon propagator. Also their longitudinal parts add
up to zero for the terms proportional to $C_F$ only. This just checks the well known
properties of the Abelian theory. It does not hold for the
$C_A$ parts of Eqs. \ref{eq:pi1t} and \ref{eq:pi4t} as they would get modified 
by the additional
diagrams. These, however, were calculated in this work without the above
reduction scheme as follows:

We use the result of the integrated fermion loop which reads (omitting color
and coupling constant factors) \cite{pokorski}:

\begin{equation}
\pi_{\mu,\nu} (k, m) \equiv \mu^\epsilon \int \frac{d^nl}{(2 \pi)^n} \frac{ Tr 
\{\gamma_\mu 
\left( \rlap/
l -\rlap/ k + m \right) \gamma_\nu \left( \rlap/ l + m \right) \} }
{(l^2-m^2)((l-k)^2-m^2)} \equiv \left( k^2 g_{\mu,\nu}-k_\mu k_\nu \right) \pi
(k^2 ,m^2) \label{eq:pidef}
\end{equation}

with

\begin{equation}
\pi (k^2,m^2) = \frac{i (-)^\frac{n}{2} 
\eta^\frac{\epsilon}{2}}{(4 \pi)^2} \Gamma \left( \frac{\epsilon}{2}
\right) \int^1_0 dx \frac{8 x (1-x)}{\left( \frac{k^2}{m^2} x(1-x)-1 \right)^\frac{\epsilon}{2}} \label{eq:pires}
\end{equation}

where $\eta$ is given in Eq. \ref{eq:paramdef}. 
For completeness, we also list the sum of the gluon and ghost contributions
in the Feynman gauge \cite{ryder,Muta} to the gluon propagator:

\begin{equation}
\pi_{\mu,\nu} (q) 
\equiv \left( q^2 g_{\mu,\nu}-q_\mu q_\nu \right) \pi
(q^2) \label{eq:CApidef}
= \left( q^2 g_{\mu,\nu}-q_\mu q_\nu \right) \frac{i  
\zeta^\frac{\epsilon}{2}\Gamma \left( \frac{\epsilon}{2}
\right) \Gamma \left( 2- \frac{\epsilon}{2} \right)^2 \left( 5 - \frac{
3 \epsilon}{2} \right)}{8 \pi^2 \Gamma \left(4-\epsilon \right) \left( 1-
\frac{\epsilon}{2} \right)} \label{eq:glpghpi}
\end{equation}

where $\zeta$ is given in Eq. \ref{eq:paramdef}. Now we get the following result for ${\cal M}_{gse_2}$:

\begin{eqnarray}
\Pi^2_t &=& \!\! \frac{-i g^4 C_A T_F \; \mu^\epsilon}{2(n-1)} \! \int \! \frac{d^nk 
\; \pi (k^2,m^2)}{(2\pi)^n} \! \left\{
\frac{3n-\frac{7}{2}}{(k+q)^2}+\frac{n-1-\frac{1}{n}}{k^2} + q^2 \frac{3n-
\frac{7}{2}}{(k+q)^2k^2} + \left( n-\frac{3}{2} \right) \left( \frac{1}{q^2}
-\frac{k^2}{q^2(k+q)^2}+ \right. \right. \nonumber \\
&& \!\! \left. \left. q^2\frac{k^2+2kq}{(k+q)^2k^4} \right) \! \right\} 
\label{eq:pi2t} \\
\Pi^2_l &=& \!\! \frac{-i g^4 C_A T_F \, \mu^\epsilon}{2} \! \int \! \frac{d^nk \; \pi (k^2,m^2)
}{(2\pi)^n} \!
\left\{ \! \frac{n+\frac{1}{n}
-\frac{3}{2}}{k^2} + \frac{1}{(k+q)^2} +\frac{\frac{3}{2}-n}{q^2} -\frac{
q^2}{2(k+q)^2k^2} + \! \left( \! n - \frac{3}{2} \right) \! \frac{k^2}{q^2 (k+q)^2}
\! \right\} \label{eq:pi2l}
\end{eqnarray}

and similarly for ${\cal M}_{gse_3}$:

\begin{eqnarray}
\Pi^3_t &=& \frac{-i g^4 C_A T_F \;\mu^\epsilon}{2(n-1)} \int \frac{d^nk}{(2\pi)^n} \pi (k^2,m^2) \left\{
\frac{6n - 2 n^2 + \frac{2}{n} -6}{k^2} \right\} \label{eq:pi3t} \\
\Pi^3_l &=& \frac{-i g^4 C_A T_F \; \mu^\epsilon}{2}\int \frac{d^nk}{(2\pi)^n} \pi (
k^2,m^2) \left\{ 
\frac{4-2n - \frac{2}{n}}{k^2} \right\} \label{eq:pi3l} 
\end{eqnarray}

All the necessary integrals are given in appendix \ref{sec:tli}.
For the vertex correction graphs we arrive at the following representations:

\begin{eqnarray}
{\cal M}_{vc_1} &=& \frac{i g^6 C_F C_A T_F \; \eta^\epsilon}{2 (4 \pi)^4 q^2} \int^1_0 dx \int^1_0 dy 
\int^1_0 du \int^1_0 dv \;\; x \; u \left[ - \frac{B (1-u)^{\frac{\epsilon}{2}-1}
\Gamma \left( \epsilon \right)}
{2 \alpha^{2+\frac{\epsilon}{2}} \left( \frac{-q^2}{m^2} \left( - \frac{\sigma^2}
{\alpha^2} + \frac{\rho}{\alpha} \right) + \frac{1-u}{\alpha} \right)^\epsilon}
\right. \nonumber \\
&& - \left. \frac{(a+nb) (1-u)^\frac{\epsilon}{2} \Gamma \left( \epsilon \right)}
{2 \alpha^{3+\frac{\epsilon}{2}} \left( \frac{-q^2}{m^2} \left( - \frac{\sigma^2
}{\alpha^2} + \frac{\rho}{\alpha} \right) + \frac{1-u}{\alpha} \right)^\epsilon}
+\frac{ c (1-u)^\frac{\epsilon}{2} \Gamma \left( 1+ \epsilon \right)}
{\alpha^{3+\frac{\epsilon}{2}} \left( \frac{-q^2}{m^2} \left( - \frac{\sigma^2}
{\alpha^2} + \frac{\rho}{\alpha} \right) + \frac{1-u}{\alpha} \right)^{1+
\epsilon}} \right] \label{eq:vc1res}
\end{eqnarray}

where $\sigma$ is given by Eq. \ref{eq:sigma}, $\rho$ by Eq. \ref{eq:rho}
and $\alpha$ by Eq. \ref{eq:paramdef}. The remaining abbreviations read:

\begin{eqnarray}
B &\equiv& -24(1-x)+8+\epsilon (12 (1-x)-4) \label{eq:vc1A} \\
a &\equiv& -16 x (1-x)^2 \label{eq:vc1a} \\
b &\equiv& 12 x (1-x)^2 \label{eq:vc1b} \\
c &\equiv& (8 - 12 x) +\frac{q^2}{m^2} \left[ \frac{\sigma^2}{\alpha^2} 12 x (1-x)^2
+ 8 (1-x)x(1-y) -12 (1-x)x^2(1-y)^2 +4 (x^2(1-y)^2 -  \right. \nonumber \\
&& \left. x(1-y)) - 2 \left( 12 (1-y)x(1-x)^2 - 8 (1-x)x(1-y) +4 x(1-x) \right)
\frac{\sigma}{\alpha} \right] \label{eq:vc1c}
\end{eqnarray}

The HQET Feynman rules of Fig. \ref{fig:hqeFr} project all three Lorentz
indices to zero for ${\cal M}_{vc_1}$. The completely antisymmetric nature
of the three gluon vertex then implies that there is no divergence coming
out of the internal fermion loop. Although Eq. \ref{eq:vc1res} appears 
to possess a double pole, the $\frac{1}{1-u}$ `` divergence"
is actually finite when integrated over all Feynman parameters. We checked this
directly with VEGAS \cite{Lepage} and it indeed gives a well converged numerical answer.
As for ${\cal M}_{vc_2}$, we integrate out the fermion loop as before, which yields:

\begin{eqnarray}
{\cal M}_{vc_2} &=& 
\frac{-i g^6 C_F C_A T_F}{2 q^2} 
\int \frac{d^nk}{(4 \pi)^n} 
\pi (k^2,m^2) \left[- \frac{k^2+2kq}{(k+q)^2k^4} 
+ \right. \nonumber \\
&& \left. \frac{1}{2(n-1)} \left( \frac{1}{(k+q)^2k^2} + \frac{1}{2} \left(
\frac{k^2+2kq}{(k+q)^2 k^4} + \frac{1}{q^2k^2} - \frac{1}{q^2(k+q)^2} 
\right) \right) \right] \label{eq:vc2res}
\end{eqnarray}

All the integrals left are given in appendix \ref{sec:tli}.

\subsection{Infra-Red Cancellations} \label{sec:IR}

In this section we turn to diagrams which give integrals already present in
an Abelian theory, however, which do not contribute in QED due to a 
cancellation that fails in the case of QCD. The reason is as follows: The color
factors for the ladder diagrams are proportional to $C_F^2$ for the straight 
and $C_F^2-\frac{C_AC_F}{2}$ for the crossed ladder graph. The same
structure is also present in graphs Eq. \ref{eq:vc3def} and \ref{eq:olvcdef}.
In the sum
of all four occurring ladder diagrams with one fermion loop plus 
${\cal M}_{vc_3}$ and ${\cal M}_{olvc}$, all terms proportional
to $C_F^2$ give a contribution that is equal to the product of the one loop
fermion graph with the Born contribution. This is an explicit example of the
aforementioned exponentiation that occurs on the level of the potential.
In an Abelian theory one thus has to omit these contributions.

On the other hand, in QCD, we need to calculate the crossed ladder terms and
keep only the $-\frac{C_AC_F}{2}$ part of the above color factors. 

From direct inspection it is furthermore obvious that these diagrams contain
infra-red (IR) divergent terms which have to cancel in the potential. It has been
shown in Refs.
\cite{Fischler77,Peter97} that the sum of 
${\cal M}_{cl}+{\cal M}_{vc_3}+{\cal M}_{olvc}$ is
IR-finite. This requirement
poses a strong check on the calculation and necessitates the calculation of
the IR-divergent parts of a diagram that vanishes in dimensional 
regularization (${\cal M}_{olvc}$), i.e. when UV- and IR-divergences are not separated.

The presence of the square of the heavy quark propagator complicates the 
calculation of the crossed ladder diagram considerably as it makes the 
analytical separation of the double and single pole terms extremely difficult.
We therefore found it most convenient to introduce a gluon mass $\lambda$ 
as an IR-regulator. This allows to explicitly differentiate between UV- and
IR-divergences and provides a strong numerical check on the sum of all 
IR-divergent contributions. In this case we get the following integral 
representations for the unrenormalized and IR-regulated amplitudes:

\begin{eqnarray}
{\cal M}_{cl} &=& 
\frac{i g^6 C_F C_A T_F}{2} 
\int \frac{d^nk}{(4 \pi)^n} \frac{\pi(k^2,m^2) (k^2-k_0^2)}{(k_0+i \varepsilon)^2
(k^2-\lambda^2+i \varepsilon)^2 ((k+q)^2-\lambda^2+i \varepsilon)} \label{eq:cl} \\
{\cal M}_{vc_3} &=& 
\frac{-i g^6 C_F C_A T_F}{2 q^2} 
\int \frac{d^nk}{(4 \pi)^n} \frac{\pi(k^2,m^2) (k^2-k_0^2)}{(k_0+i \varepsilon)^2
(k^2-\lambda^2+i \varepsilon)^2} \label{eq:vc3} \\
{\cal M}_{olvc} &=& 
\frac{-i g^6 C_F C_A T_F }{2 q^4} \pi(q^2,m^2)
\int \frac{d^nk}{(4 \pi)^n} \frac{1}{(k_0+i \varepsilon)^2
(k^2-\lambda^2+i \varepsilon)} \label{eq:olvc} 
\end{eqnarray}

For the contributions of graphs \ref{eq:cl} and \ref{eq:vc3} in which the 
$k_0^2$ terms in
the numerator cancel the heavy quark propagator, no gluon mass regulator 
in needed. The sum of these $k_0$-independent parts of ${\cal M}_{cl}$ and
${\cal M}_{vc_3}$ are separately IR-finite and indeed proportional to the
integral \ref{eq:I5} in appendix \ref{sec:tli}. We therefore restrict ourselves
to a discussion of the $k_0$-dependent contributions only.
In these integrals the $i$-$\varepsilon$ prescription is crucial in order to
arrive at the correct location of poles and branch cuts the in the complex 
$k_0-plane$. The presence of the fermion
mass brings about a complicated integral over a general power, which in turn
leads to a branch cut in the upper half of the plane. After integrating over
$k_0$ in such a fashion one is left with an Euclidean integral over 
$(n-1)$-dimensions. More details and complete results are given in appendix \ref{sec:tlgm}.

\section{Renormalization} \label{sec:ren}

In Fig. \ref{fig:hqpct} we list the relevant counterterms for the 
two loop diagrams of
Fig. \ref{fig:hqpfig1} and Fig. \ref{fig:hqpfig1a}. 
The counterterms themselves 
contain non-local contributions, i.e. non-polynomial in the momentum transfer $q$,
that have to cancel the non-local terms from the original amplitudes. The 
construction of the local wave function renormalization constants 
provides a powerful test of the correctness of the results presented
both in section \ref{sec:res} and the appendices as they must combine 
successfully to arrive at the required local double and single pole terms. 
It might be helpful to expound on the general treatment of masses within the
corresponding integrals and counterterms in
the MS-renormalization scheme \cite{Collins,PeskSchr}. In the counterterm
approach, their contribution is restricted to finite changes
through the counterterms as the wave function renormalization constants are
independent of the fermion masses. In other words, all pole terms that contain
masses represent non-local infinities which must cancel in the sum of graphs
contributing to the overall field strength renormalization. There is therefore
no difference in the formal treatment of the mass parameter in graph \ref{eq:gse5def}
and any other graph. This is another way of saying that the parameters of a
MS-renormalized theory are not physical. Rather, they are related to measurable
quatities by a perturbative series in the physical parameters.

We begin by presenting the results for the counterterms corresponding to
Fig. \ref{fig:hqpct}. All two point counterterms correspond to the transverse
parts of the gluon self energy contribution only, as these are the only
relevant ones for this work. The graph ${\cal M}_{gse_1}$ has two counterterms, one stemming
from the fermion loop divergence ($\Pi_{ct_{1f}}$) and one from the loop around
the three gluon vertex ($\Pi_{ct_{1g}}$). They are given in the MS-renormalization
scheme:

\begin{center}
\begin{figure}
\centering
 \epsfig{file=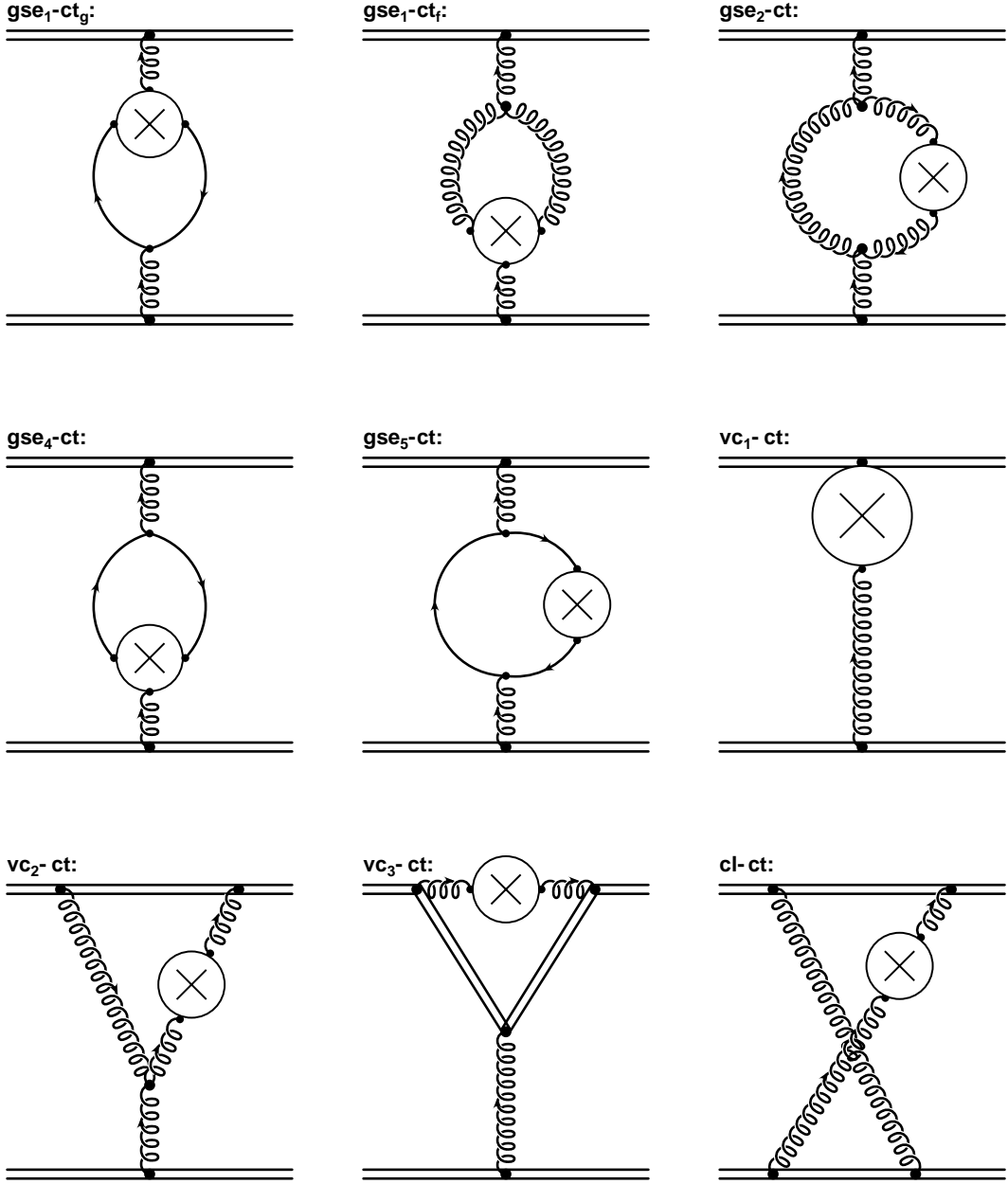,width=14cm}
 \caption{The two loop counterterms corresponding to the diagrams in Figs. 
 \ref{fig:hqpfig1} and \ref{fig:hqpfig1a}.
Adding these contributions to the original graphs removes all non-local 
functions from the occurring pole terms. The only exception are $\frac{m^2}{
\epsilon}$ terms in the two point functions which only cancel in the sum
of all two point diagrams as explained in the text. The fact that the tadpole
diagram has no counterterm is already indicative of this cancellation.}
 \label{fig:hqpct}
\end{figure}
\end{center}

\begin{eqnarray} 
\Pi_{ct_{1_f}} &=&  \frac{-8i g^4 C_A T_F
}{3(4 \pi)^4 \epsilon} \int^1_0 \!\! dx \! \left[ \left( 8
- 6n \right) \frac{n}{2} \left( \frac{n}{2} m^2 \Gamma \! \left( -1 + \frac{\epsilon
}{2} \right) \!\! \left( -\frac{q^2}{m^2} x (1-x) \right)^{1-\frac{\epsilon}{2}}
\!\!\!\!\!\! -q^2 x^2 \Gamma \!
\left( \frac{\epsilon}{2} \right) \!\! \left( -\frac{q^2}{m^2} x (1-x) \right)^{-\frac{\epsilon}{2}}
\right) \right. \nonumber \\
&& - \left( 4 n -6 \right) \left( \frac{m^2}{2} \Gamma \! \left( -1 + \frac{\epsilon}{2} 
\right) \!\! \left( -\frac{q^2}{m^2} x (1-x) \right)^{1-\frac{\epsilon}{2}} 
\!\!\!\!\!\! -q^2 x^2 \Gamma \! 
\left( \frac{\epsilon}{2} \right) \!\! \left( -\frac{q^2}{m^2} x (1-x) \right)^{-\frac{\epsilon}{2}}
\right) +  \left( q^2  \left( -2 + 2n \right) x \right. \nonumber \\
&&  \left. \left. + q^2 \left( 5 n -5 \right) \Gamma \!
\left( \frac{\epsilon}{2} \right) \!\! \left( -\frac{q^2}{m^2} x (1-x) \right)^{-\frac{\epsilon}
{2}} \right) \right] \frac{\eta^\frac{\epsilon}{2}}{n-1} \label{eq:gse1ctf} \\
\Pi_{ct_{1_g}} &=&  \frac{6i g^4 C_A T_F}{(4 \pi)^4 \epsilon} \int^1_0 \!\! 
dx \left[ \left( 4 n -12 \right)
\left( - \frac{n}{2} m^2 \Gamma \! \left( -1 + \frac{\epsilon}{2} \right) \!\! \left(
-\frac{q^2}{m^2} x(1-x) + 1 \right)^{1-\frac{\epsilon}{2}} \!\!\!\!\!\! + q^2 x^2 \Gamma \! \left(
\frac{\epsilon}{2} \right) \!\! \left(-\frac{q^2}{m^2} x(1-x) + \right. \right. \right. \nonumber \\ 
&& \left. \left. 1 \right)^{-\frac{\epsilon}{2}} \right)
- ( 4n -4 ) ( m^2 + q^2 x ) \Gamma \! \left( \frac{ \epsilon}{2} \right) \!\!
\left( -\frac{q^2}{m^2} x(1-x) + 1 \right)^{-\frac{\epsilon}{2}} \!\!\!\! - 4 m^2 \Gamma \! \left( -1 +
\frac{\epsilon}{2} \right) \!\! \left( -\frac{q^2}{m^2} x(1-x) + 1 \right)^{1-\frac{\epsilon}{2}} \nonumber \\
&& \left. -q^2x^2 \Gamma \! \left(
\frac{\epsilon}{2} \right) \!\! \left( -\frac{q^2}{m^2} x(1-x) + 1 \right)^{-\frac{\epsilon}{2}}
 \right] \frac{\eta^\frac{\epsilon}{2}}{n-1} \label{eq:gse1ctg}
\end{eqnarray}

For the counterterm for ${\cal M}_{gse_2}$ we find:

\begin{equation}
\Pi_{ct_2} =  \frac{4i g^4 C_A T_F \; q^2 \zeta^\frac{\epsilon}{2}}{3(4 \pi)^4 \epsilon  (n-1)} \left(
\left( \frac{7}{2}-3n \right) \frac{\Gamma \left( \frac{\epsilon}{2} \right)
\Gamma^2 \left( 1 - \frac{\epsilon}{2} \right)}{\Gamma \left( 2 - \epsilon
\right)} + \left( n - \frac{3}{2} \right) \frac{ \Gamma \left( 1+ \frac{\epsilon}{2} \right) 
\Gamma \left( 1- \frac{ \epsilon}{2} \right) \Gamma \left( - \frac{\epsilon}{2} \right)}
{\Gamma \left( 1- \epsilon \right)} \right) \label{eq:gse2ct}
\end{equation}

where $\eta$ and $\zeta$ are defined in appendix \ref{sec:tli}. For ${\cal M}_{gse_3}$ there is no counterterm
as the subdivergence is independent of the mass which means that in dimensional
regularization all the remaining integrals vanish.

The pole terms for the respective terms, expanded up to 
${\cal O} \left( \epsilon^0 \right)$, thus read:

\begin{eqnarray}
\left[ \Pi^1_t + \Pi_{ct_{1_f}} + \Pi_{ct_{1_g}} \right]_{{\cal O} \left( \epsilon^0 \right)}
&=& - \frac{i g^4 C_A T_F \; q^2}{(4 \pi)^4} \left( \frac{1}{9 \epsilon^2} + \frac{163}{108 \epsilon} - 
\frac{3m^2}{q^2 \epsilon} \right) \label{eq:gse1ren} \\
\left[ \Pi^2_t + \Pi_{ct_2} \right]_{{\cal O} \left( \epsilon^0 \right)}
&=& - \frac{i g^4 C_A T_F \; q^2}{(4 \pi)^4} \left( - \frac{44}{9 \epsilon^2} + \frac{25}{27 \epsilon} + 
\frac{15m^2}{q^2 \epsilon} \right) \label{eq:gse3ren} \\
\left[ \Pi^3_t \right]_{{\cal O} \left( \epsilon^0 \right)} &=&  \frac{i g^4 
C_A T_F \; 18 m^2}{(4 \pi)^4 \epsilon}
\end{eqnarray}

These equations contain no non-local terms other than the $\frac{m^2}{\epsilon}$
terms, which then have to vanish in the sum of all contributions to the
non-Abelian part of the gluon wave function renormalization constant. 
Because of the very involved nature of the occurring non-local terms, this
is already powerful evidence of the correct evaluation of both the
two loop integrals as well as the decomposition of graph ${\cal M}_{gse_1}$.
Multiplying each graph with its respective multiplicity 
we find in the
MS-scheme:

\begin{equation}
\left\{ 4 \left[ \Pi^1_t + \Pi_{ct_{1_f}} + \Pi_{ct_{1_g}} \right] + 2 \left[ \Pi^2_t + \Pi_{ct_2} \right]
+ \left[ \Pi^3_t \right] \right\}_{{\cal O} \left( \epsilon^0 \right)} = 
 \frac{i g^4 C_A T_F \; q^2}{(4 \pi)^4} \left(  \frac{28}{3 \epsilon^2} - \frac{71}{9 \epsilon} \right) \label{eq:nArc}  
\end{equation}

This is completely local and thus demonstrates that the renormalization has
been carried out properly and that the integrals given are correct. In order to 
further check this term we need 
the pole term from the ``overlapping'' Abelian two point diagram from Eq. \ref{eq:gse4def} (which in QCD
develops a color factor proportional to ($C_F- \frac{1}{2} C_A$) in order to
get the fermionic part of the overall gluon wave function renormalization constant $Z_3$.
The counterterm for ${\cal M}_{gse_4}$ reads

\begin{equation}
\Pi_{ct_4} = - \frac{8 \;i \;g^4}{\epsilon(4 \pi)^2} \left(C_F- \frac{C_A}{2} \right) T_F \; q^2 
\;\; \pi \left( q^2, m^2
\right) \label{eq:gse4ct}
\end{equation}

and gives in agreement with \cite{itzzub}: 

\begin{equation}
\left\{ \left[ \Pi^4_t + \Pi_{ct_4} \right]
\right\}_{{\cal O} \left( \epsilon^0 \right)} = 
 \frac{i g^4 (C_F-\frac{C_A}{2}) T_F \; q^2}{(4 \pi)^4} \left(  \frac{16}{3 \epsilon^2} - \frac{52}{9 \epsilon} \right) \label{eq:Arc}  
\end{equation}

Adding Eqs. \ref{eq:nArc} and  the $C_A$ term of \ref{eq:Arc} gives the correct 
non-Abelian fermionic part of
the gluon wave function renormalization constant ((times $\frac{1}{i q^2}$) 
see Ref. \cite{Muta} for example)
in the Feynman gauge:

\begin{equation} 
Z^{C_A}_{3_{fermionic}} = 
 \frac{ g^4 C_A T_F}{(4 \pi)^4} \left(  \frac{20}{3 \epsilon^2} - \frac{5}{
  \epsilon} \right) \label{eq:fZ3}
\end{equation}

This testifies to the overall correctness of both the decompositions used
as well as all the integrals listed in the appendices!

For completeness we also give the counterterm for ${\cal M}_{gse_5}$, which 
in the MS-scheme must 
be treated in the same way as the graphs before. All divergent terms proportional
to $m^2$ cancel the corresponding non-local infinities in Eq. \ref{eq:pi5t}:

\begin{eqnarray}
\Pi_{ct_5} &=& - \frac{4 \;i \;g^4 C_F T_F}{\epsilon (n-1) (4 \pi)^2} \left[
n \left( 12 m^2 B_{22} +2q^2B_{12}- 4 A_2 - 12 q^2 m^2 C_{122} \right)
+(24q^2m^2-48 m^4) C_{122} \right.
\nonumber \\ && \left. -24 m^2 B_{22} -(4q^2+16 m^2) B_{12} +8 A_2 \right]
\label{eq:gse5ct}
\end{eqnarray}

with 
 
\begin{equation}
\left\{ \left[ \Pi^5_t + \Pi_{ct_5} \right]
\right\}_{{\cal O} \left( \epsilon^0 \right)} = 
 \frac{i g^4 C_F T_F \; q^2}{(4 \pi)^4} \left(  - \frac{8}{3 \epsilon^2} + \frac{8}{9 \epsilon} \right) \label{eq:CFArc}  
\end{equation}

It is an important difference to the massless case that the counterterms \ref{eq:gse4ct} (rather its $C_F$ part)
and \ref{eq:gse5ct} are not related by a simple minus sign as implied by the Ward
identity. There is an additional constant term $4 m$ which gives new contributions.
For the purely Abelian fermionic part of the gluon wave function renormalization
constant in the Feynman gauge we find in agreement with Ref. \cite{Muta}:

\begin{equation} 
Z^{C_F}_{3_{fermionic}} = 
 \frac{ g^4 C_F T_F}{(4 \pi)^4} \left( - \frac{4}{
 \epsilon} \right) \label{eq:fZ3CF}
\end{equation}

The cancellation of the higher order (double) pole is a characteristic feature in QED that
holds to all orders \cite{Adler}.

For ${\cal M}_{vc_1}$ we do not need to remove non-local terms as the fermion loop is finite
due to the projection of all three Lorentz indices to zero. It is easy to check
this by calculating all divergent pieces after the integration of the fermion loop.
All that is left is the divergence from the remaining integral  which has to 
be subtracted in the usual MS-fashion. This is indicated in Fig. \ref{fig:hqpct}.
The explicit pole term is given by:

\begin{equation}
\left[ {\cal M}_{vc1} \right]_{{\cal O} \left( \epsilon^0 \right)} = \frac{ig^6 C_F C_A T_F
}{(4 \pi)^4 q^2} \left( - \frac{1}{\epsilon} \right) 
\end{equation}

in agreement with the massless case \cite{Peter}. In the case of ${\cal M}_{vc_2}$ we do have non-local terms, and the counterterm reads:

\begin{eqnarray}
{\cal M}_{vc2_{ct}} &=& \frac{4i g^6 C_F C_A T_F \; \eta^\frac{\epsilon}{2}}{3
(4 \pi)^4q^2 \epsilon} \int^1_0 dv \left( - \frac{
n v \Gamma \left( \frac{\epsilon}{2} \right) }{2 (\frac{-q^2}{m^2} v (1-v))^\frac{\epsilon}
{2}} + \frac{ (1+v) \Gamma \left( 1 + \frac{\epsilon}{2} \right) }
{(\frac{-q^2}{m^2} v (1-v))^\frac{\epsilon}{2} } + \right. \nonumber \\
&& \left. \frac{ \Gamma \left( \frac{
\epsilon}{2} \right) }{2 (n-1) (\frac{-q^2}{m^2} v (1-v))^\frac{\epsilon}{2}}
+ \frac{ n v \Gamma \left( \frac{\epsilon}{2} \right) }{ 8 (n-1)
(\frac{-q^2}{m^2} v (1-v))^\frac{\epsilon}{2}} - \frac{ (1+v) \Gamma \left(
1+ \frac{\epsilon}{2} \right) }{4 (n-1) (\frac{-q^2}{m^2} v (1-v))^\frac{\epsilon}{2}}
\right) \label{eq:vc2ct}
\end{eqnarray}

Adding Eq. \ref{eq:vc2ct} with the appropriate normalization and color factors
to the result given in \ref{eq:vc2res} does indeed give completely local
double and single pole terms as required in dimensional regularization after
the subdivergences are subtracted: 

\begin{equation}
\left[ {\cal M}_{vc_2} + {\cal M}_{vc2_{ct}} \right]_{{\cal O} \left( \epsilon^0 \right)} = \frac{ig^6 C_F C_A T_F
}{(4 \pi)^4 q^2} \left( \frac{1}{\epsilon^2} - \frac{5}{12\epsilon}
\right)
\end{equation}

It demonstrates that 
indeed all non-local divergences 
are canceled and agrees furthermore with the pole terms obtained in the massless
analysis \cite{Peter}!
It should be noted that all the integrals needed were already used in the
${\cal M}_{gse_2}$ calculation for which such a strong internal consistency 
check was performed just above.
All the required expansions above were carried out with the help of MAPLE in
face of the complexity involved. As mentioned before, also the translation into
FORTRAN was handled by MAPLE as to reduce possible accidental errors. 
.

\subsection{Counterterms with Gluon Mass}

At this point we need the counterterms of the IR-divergent contributions,
${\cal M}_{cl}$, ${\cal M}_{vc_3}$ and ${\cal M}_{olvc}$. As indicated above
and expressed in Eqs. \ref{eq:cl}, \ref{eq:vc3} and \ref{eq:olvc}, these were regulated by
introducing a gluon mass regulator. The remaining UV-divergences are treated
as above in the context of dimensional regularization. We therefore have to
calculate all counterterm contributions that occur for gluon propagators
with a gluon mass. Without such a dimensionful quantity, only the crossed
ladder diagram would yield a counterterm in dimensional regularization.
We again use the gluon mass only for $k_0$-dependent terms as explained in
section \ref{sec:IR}. This is indicated below. 

\begin{center}
\begin{figure}[t]
\centering
 \epsfig{file=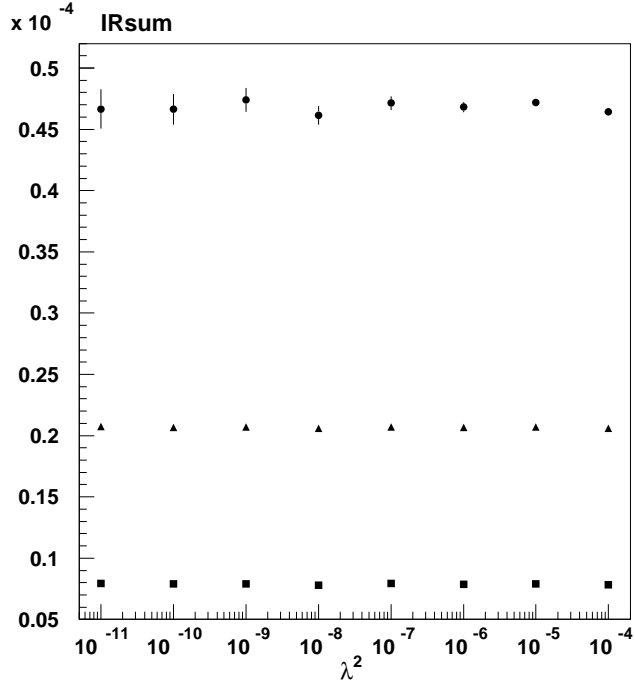,width=10cm}
 \caption{The sum of the $\lambda^2$-dependent amplitudes
 and counterterms
 ${\cal M}^{k_0}_{cl}+{\cal M}^{k_0}_{vc_3}+{\cal M}_{olvc}+
 {\cal M}^{k_0}_{cl_{ct}}+{\cal M}^{k_0}_{{vc_3}_{ct}}$. Circles
 correspond to a choice of $q^2 = -10 GeV^2$ and $m = m_c$,  
 triangles to $q^2 = -100 GeV^2$ and 
 $m = m_c$ while the lower curve (squares) has $q^2 = -100 GeV^2$ and $m = m_b$.
 The overall normalization neglects color factors and the coupling
 strength. All data are obtained by using $10^6$ evaluations per iteration
 with VEGAS and 100 iterations. The statistical error is 
 indicated and smaller than the symbols where invisible. 
 The sum for each of the displayed sets of parameters
 is clearly independent of the IR-gluon mass regulator $\lambda$ as expected.} 
 \label{fig:ldep}
\end{figure}
\end{center}

The results are obtained in a similar way as for the corresponding amplitudes,
first integrating over the heavy quark propagator in the complex $k_0$-plane
with a subsequential $(n-1)$-dimensional Euclidean integral remaining. 
The results are obtained straightforwardly as there are only pole terms and
no branch cuts in the counterterm contributions. We find for the gluon mass
regulated terms:

\begin{eqnarray}
{\cal M}^{k_0}_{cl_{ct}} &=& \frac{4i g^6 C_F C_A T_F \; \eta^\frac{\epsilon}{2} \Gamma
\left( 1 +\frac{\epsilon}{2} \right)}{ \epsilon
(4 \pi)^\frac{7}{2} \Gamma\left( \frac{5}{2} \right) m^2} \int^1_0 dv\frac{
1}{ (\frac{-q^2}{m^2} v (1-v) + \frac{\lambda^2}{m^2})^{1+\frac{\epsilon}
{2}}} \label{eq:clk0ct} \\
{\cal M}^{k_0}_{{vc_3}_{ct}} &=& - \frac{8 i g^6 C_F C_A T_F \; \eta^\frac{\epsilon}{2} 
\Gamma \left( \frac{\epsilon}{2} \right)}{3 \epsilon q^2 (4 \pi)^\frac{7}{2}
\Gamma \left( \frac{3}{2} \right) \left( \frac{\lambda^2}{m^2} \right)^
\frac{ \epsilon}{2}} \label{eq:vc3k0ct}
\end{eqnarray}

For completeness we also list the remaining counterterm stemming from the
$k_0$-independent part of ${\cal M}_{cl}$:

\begin{equation}
{\cal M}^{k}_{{cl}_{ct}} = \frac{16 i g^6 C_F C_A T_F \; 4^\frac{\epsilon}{2} 
\eta^\frac{\epsilon}
{2} \Gamma \left(1 +\frac{\epsilon}{2} \right)
\Gamma \left( -\frac{\epsilon}{2} \right)}{3 \epsilon m^2 (4 \pi)^\frac{7}{2}
\Gamma \left( \frac{1}{2} -\frac{\epsilon}{2} \right) \left( 
\frac{-q^2}{m^2} \right)^{1+
\frac{ \epsilon}{2}}} \label{eq:clkct}
\end{equation}

The gluon mass terms that occur in the expansion of the original as well as
the counterterms above 
in powers of $\epsilon$ in the pole terms of dimensional regularization 
represent now non-local divergences which have to cancel in the same way
as terms containing $m^2$ or non-polynomial functions of $q^2$. 
The remaining IR-divergent pole terms
are contained in the form of logarithmic divergences in $\lambda$.
Fig. \ref{fig:ldep} demonstrates that in the sum of the IR-divergent
amplitudes plus their corresponding counterterms no $\lambda$-dependence
is left within the statistical errors. For convenience, three sets of
values for $q^2$ and $m^2$ are displayed while the renormalization scale
$\mu$ remains fixed. We have checked that it also holds for a variety of
other choices of parameters. Some need fewer evaluations to converge
while others need up to $10^7$ per iteration.

It is perhaps interesting to note that the crossed ladder diagram, naively
only singly IR-divergent, actually possesses a quadratic divergence in
$log ( \lambda )$ which cancels the (also unexpected) quadratic divergence in 
the Abelian vertex correction term. The remaining UV-divergent pole terms
in the MS-scheme are found to be:

\begin{eqnarray}
\left[ {\cal M}^{k_0}_{cl} + {\cal M}^{k_0}_{cl_{ct}} \right]_{{\cal O} 
\left( \epsilon^0 \right)} &=& 0 \label{eq:clrc} \\
\left[ {\cal M}^{k_0}_{vc_3} + {\cal M}^{k_0}_{{vc_3}_{ct}} \right]_{{\cal O} 
\left( \epsilon^0 \right)} &=& \frac{ig^6 C_F C_A T_F
}{(4 \pi)^4 q^2} \left(- \frac{16}{3\epsilon^2} + \frac{80}{9\epsilon}
\right) \label{eq:vc3rc} \\
\left[ {\cal M}^k_{cl} + {\cal M}^k_{vc_3}  + {\cal M}^k_{cl_{ct}} 
\right]_{{\cal O} 
\left( \epsilon^0 \right)} &=& \frac{ig^6 C_F C_A T_F
}{(4 \pi)^4 q^2} \left(- \frac{16}{3\epsilon^2} + \frac{16}{9\epsilon}
\right) \label{eq:clvc3krc} 
\end{eqnarray}

which states that the counterterm in case of ${\cal M}^{k_0}_{cl}$ completely
remove all pole terms in $\epsilon$. It is also clear that all non-local
terms are removed by the appropriate counterterms as was expected.
In order to compare this with the results obtained in the massless case one would
need to differentiate between $\epsilon_{UV}$ and $\epsilon_{IR}$.

\section{Numerical Results} \label{sec:numres}

\begin{center}
\begin{figure}
\centering
 \epsfig{file=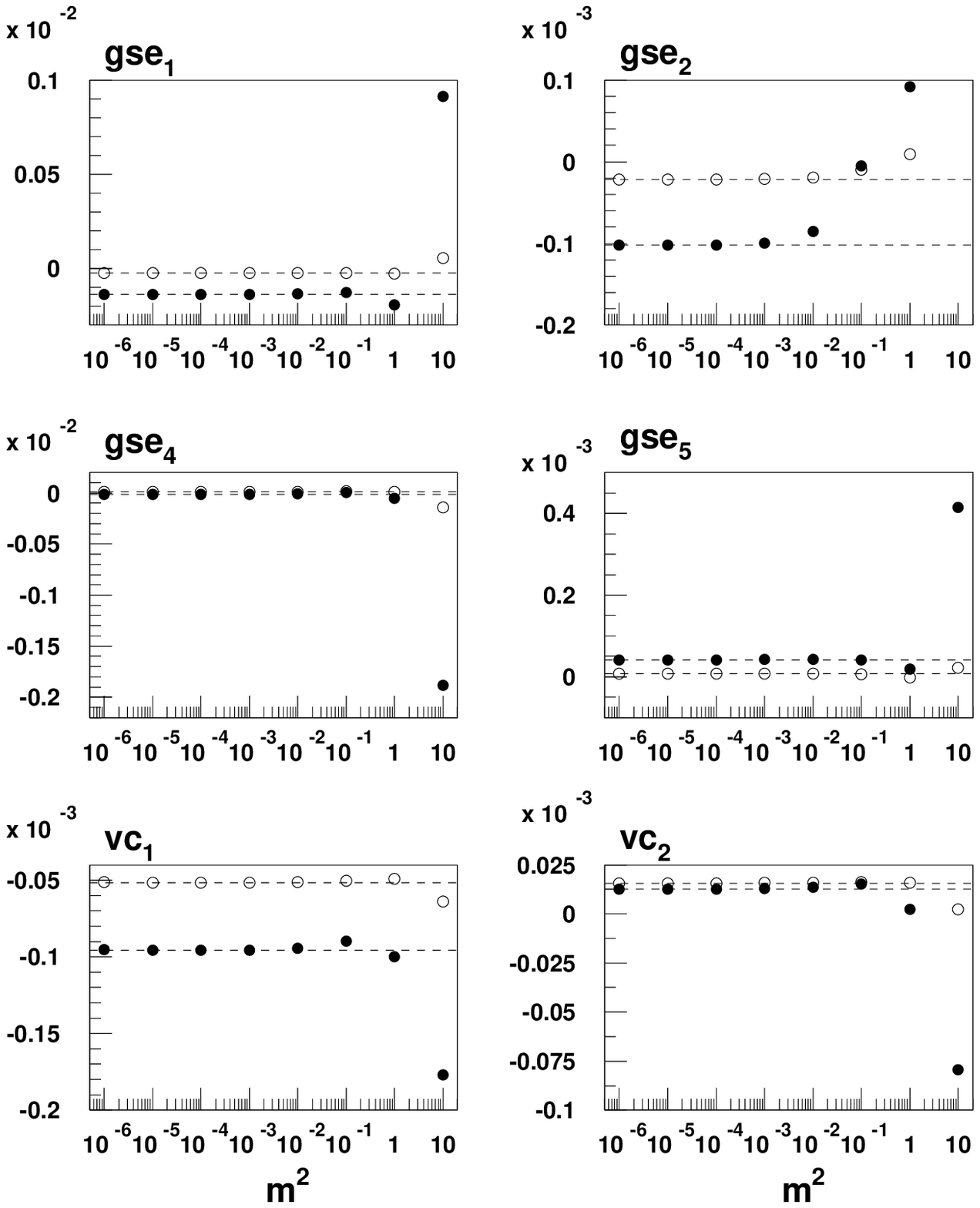,width=16cm}
 \caption{A comparison of the six amplitudes ${\cal M}_{gse_1}$, 
 ${\cal M}_{gse_2}$, ${\cal M}_{gse_4}$, ${\cal M}_{gse_5}$,
 ${\cal M}_{vc_1}$ and ${\cal M}_{vc_2}$ with the massless
 limit (dashed lines) \cite{Peter} in the $MS$-scheme. 
 Solid circles correspond to a choice of $q^2=-1.5 \;GeV^2$,
 open ones to $q^2=-4.5 \;GeV^2$. $\mu=0.31 \;GeV$ in each case. Each graph begins
 to deviate from the massless limit only when $m^2$ is of the same order as
 $-q^2$ as expected. These results were obtained after $10^6$ evaluations per
 iteration and after 50 iterations. The statistical error is smaller than the size
 of the symbols and the normalization neglects color factors and the
 coupling strength.}
 \label{fig:mcomp}
\end{figure}
\end{center}

At this point we have calculated all diagrams that contribute to the massive
fermionic corrections to the heavy quark potential that were previously
unknown. 
In the previous section we demonstrated that the counterterms 
successfully remove all non-local divergences and that the MS-subtraction terms
coincide with the massless limit. The complexity of the explicit results
given in the appendices raises some questions about how stable a numerical 
integration over up to four Feynman parameters is with VEGAS as well as about
the correctness of the finite terms of these expressions. An ideal test is
provided by the results obtained in Ref. \cite{Peter97} for the massless limit.

\begin{center}
\begin{figure}[t]
\centering
 \epsfig{file=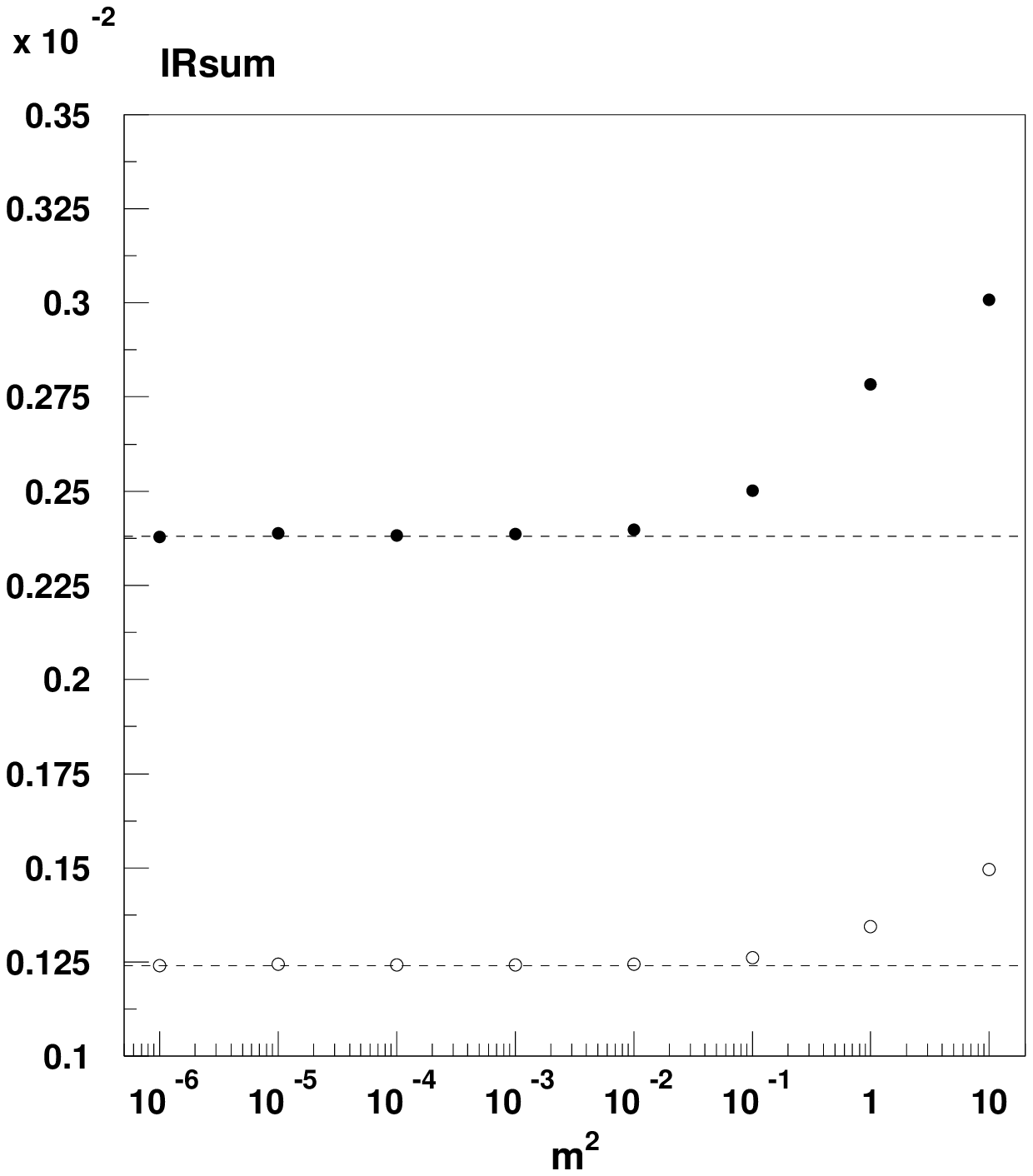,width=10cm}
 \caption{A comparison of the sum of amplitudes ${\cal M}_{cl} 
 +{\cal M}_{vc_3}+{\cal M}_{olvc}$ plus their $MS$-counterterms with the massless
 limit (dashed lines) \cite{Peter}. Solid circles correspond to a choice of $q^2=-1.5 \;GeV^2$,
 open ones to $q^2=-4.5\; GeV^2$. $\mu=0.031 \;GeV$ and $\lambda^2=10^{-8}$ in each case.
 The sum begins to deviate from the massless limit only when $m^2$ is of the same 
 order as $-q^2$ as was the case for the other graphs. These results were obtained
 after $10^6$ evaluations per
 iteration and after 100 iterations. 
 The statistical error is smaller than the size
  of the symbols and 
 the normalization neglects color factors and the
 coupling strength.}
 \label{fig:lmcomp}
\end{figure}
\end{center}

Fig. \ref{fig:mcomp} contains the results of the IR-finite two loop 
amplitudes from Figs. \ref{fig:hqpfig1} and \ref{fig:hqpfig1a} 
in section \ref{sec:nac}. The tadpole
diagram vanishes trivially in that limit so that only the six graphs shown
remain. The sames choices for $q^2$ and the renormalization scale $\mu$ were
made in all six plots. Since the results of Ref. \cite{Peter97} were calculated
in the $\overline{MS}$-renormalization scheme, we use

\begin{equation}
\mu_{_{MS}}= \sqrt{\frac{e^\gamma}{4 \pi}} \mu_{_{\overline{MS}}}
\end{equation}

It is clear from these results that deviations from the massless limit only occur
when $m^2 \approx -q^2$ or greater. This was of course expected and the motivation for this
calculation. A similar dependence is observed for the sum of the three IR-divergent
amplitudes from Fig. \ref{fig:hqpfig1} in section \ref{sec:nac}. Here it is 
impossible to compare on an amplitude by amplitude level since a different 
IR-regulator was used. 
Only the sum of infra-red finite contributions can be compared
at the two loop level. We checked explicitly that by replacing $log( \lambda)$
with $\frac{1}{\epsilon}$, only the double pole terms can be seen to be identical.

\begin{center}
\begin{figure}
\centering
 \epsfig{file=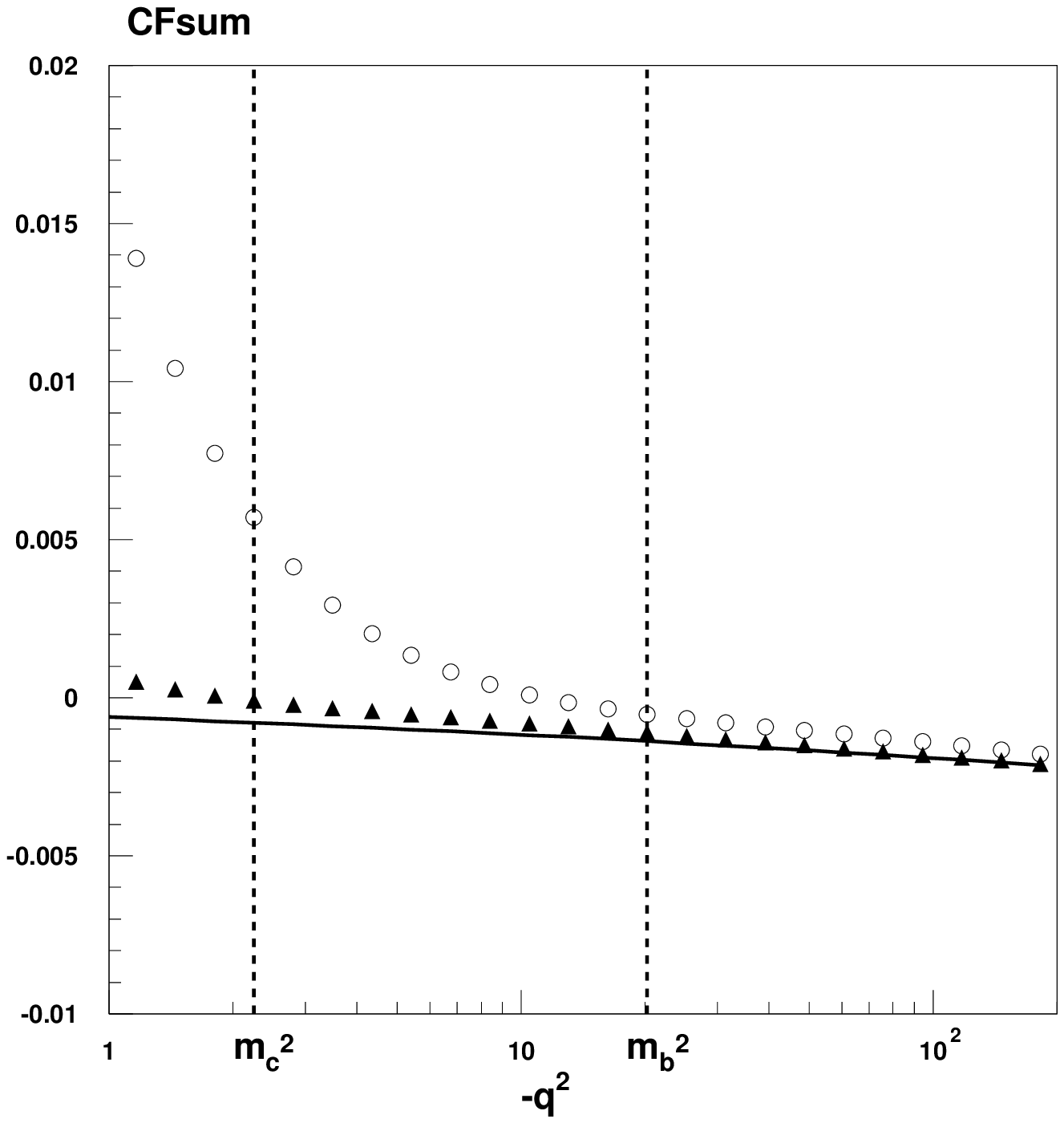,width=8cm}
 \epsfig{file=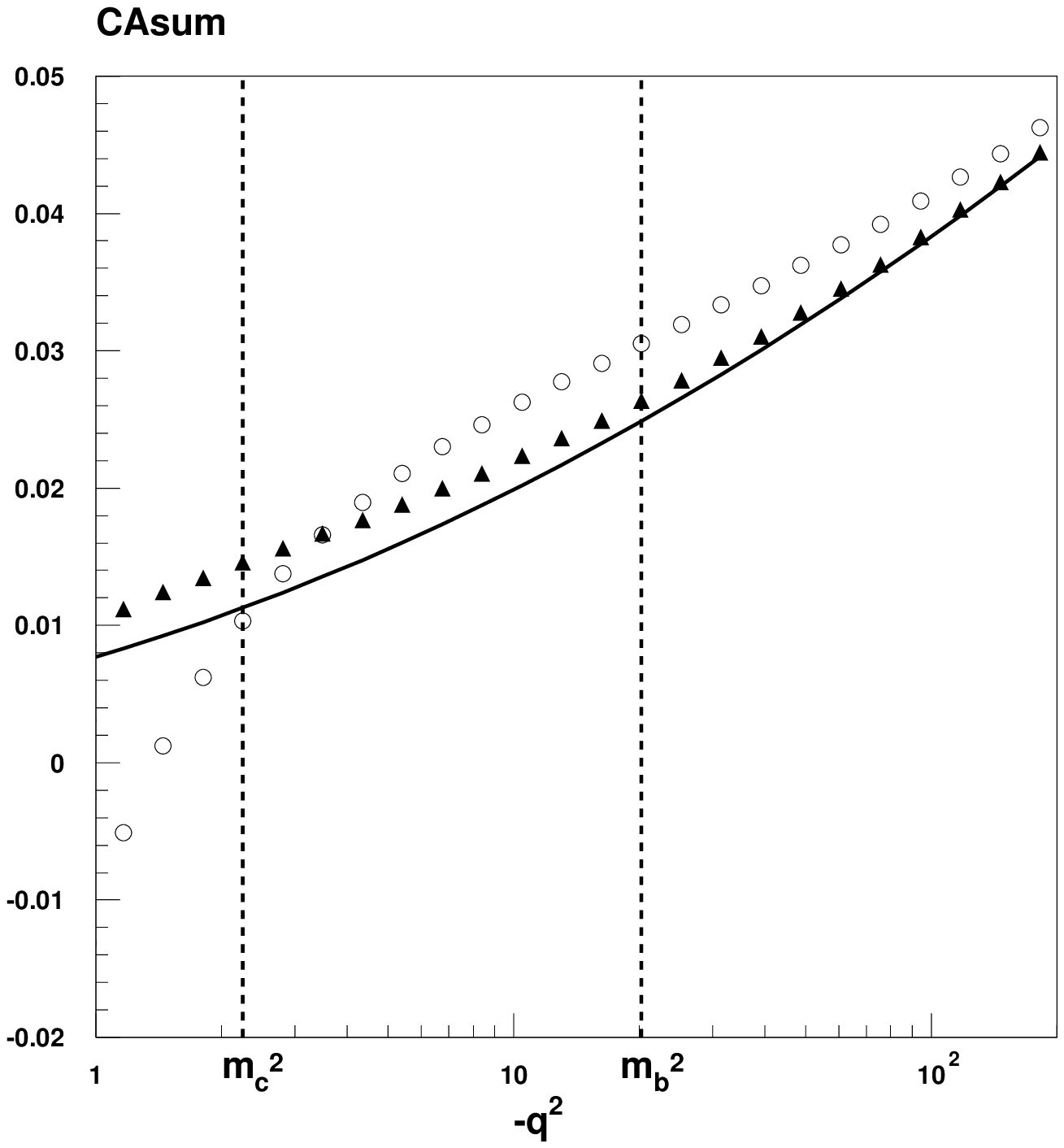,width=8cm}
 \epsfig{file=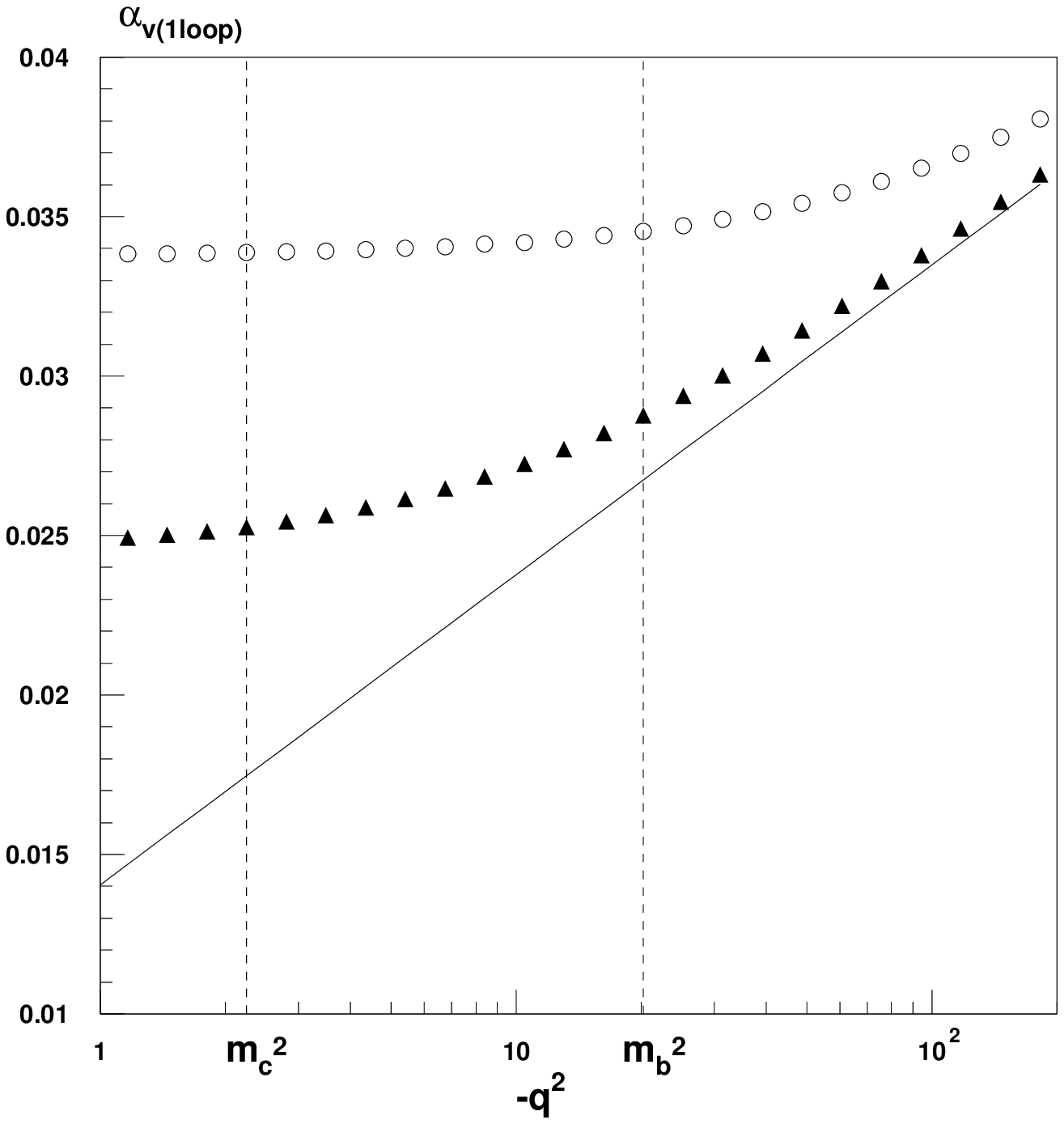,width=8cm}
 \caption{The mass dependence of $ \alpha_V$ at one (bottom) and two loops.
 The two loop case is displayed in terms of 
 all Abelian terms (left)
 and for all non-Abelian terms (proportional to $C_A$).
 Triangles denote $m^2=m_c^2=(1.5 GeV)^2$ and open circles $m^2=m_b^2
 =(4.5 GeV)^2$. The massless case is also included (line). All curves have
 the same value of the renormalization scale $\mu = 0.031$. It is clearly visible
 that the flavor threshold behavior is quite similar in the three figures with 
 an 
 opposite tendency for low values of $-q^2$ in the two loop case though. 
 The one loop corrections
 have an opposite sign relative to the Abelian two loop corrections. 
 The coupling
 constants are omitted. All cases
 approach the massless limit when $\frac{m^2}{-q^2} \ll 1$.}
 \label{fig:caqcfq}
\end{figure}
\end{center}

\begin{center}
\begin{figure}[t]
\centering
 \epsfig{file=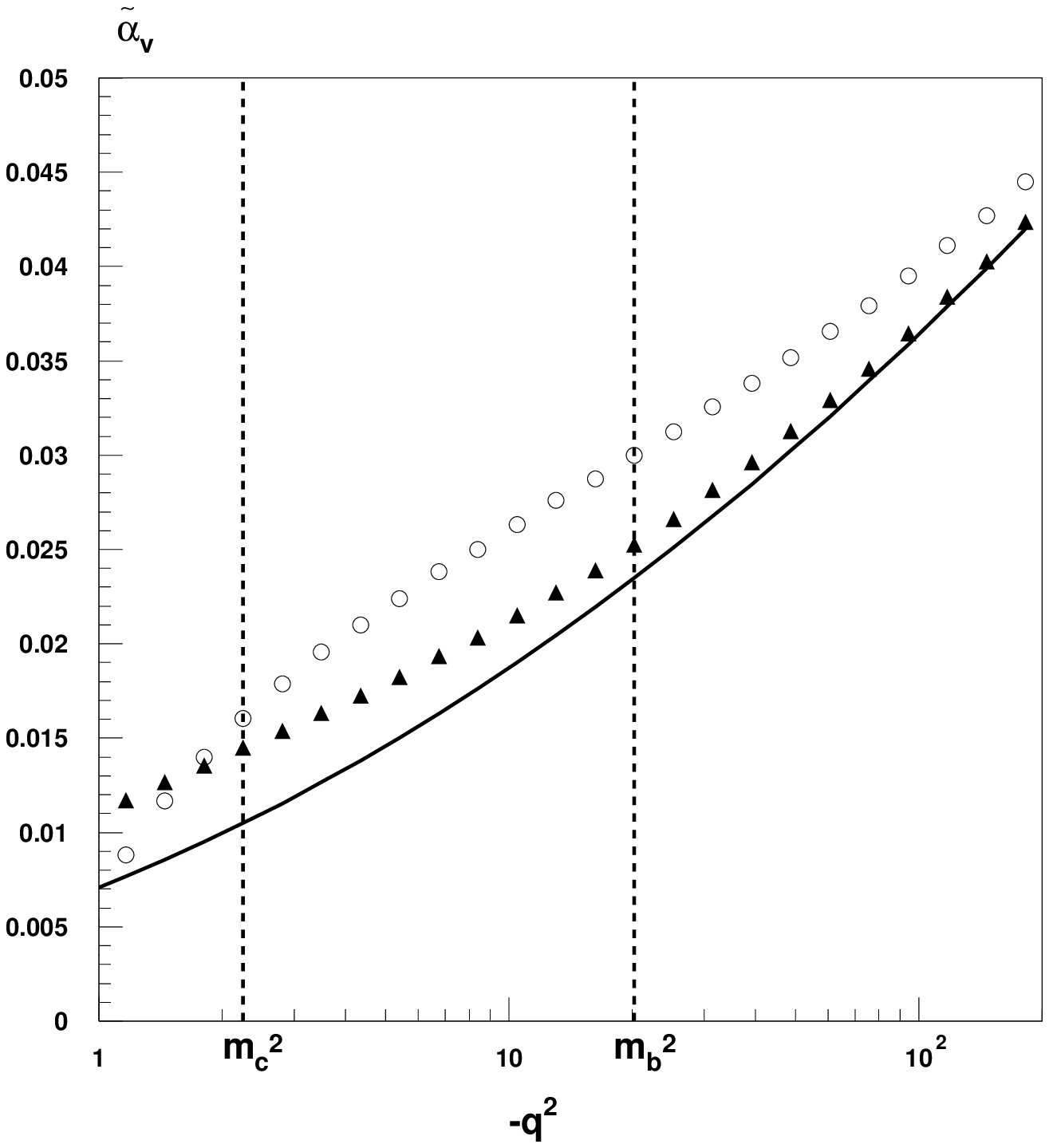,width=10cm}
 \caption{The complete two loop mass dependence of $\tilde{\alpha_V} \equiv \frac{
 \alpha^{f_{2\,loop}}_V}{g^6}$ for $m^2=m_c^2=(1.5 GeV)^2$ (triangles) and $m^2=m_b^2
 =(4.5 GeV)^2$ (open circles). The massless case is also included (line).
 In all three curves we use $\mu=0.031$. The deviation from the massless case
 at the flavor thresholds is of order of $33\%$ and is dominated by the new
 non-Abelian contributions} 
 \label{fig:cacf}
\end{figure}
\end{center}

The single pole terms differ and so do the finite contributions for each amplitude.
In the sum, however, the IR-divergent pieces cancel (as demonstrated in Fig. 
\ref{fig:ldep}), and here we can find a meaningful comparison. Fig. \ref{fig:lmcomp}
demonstrates that the correct massless limit is indeed recovered. The numerical
accuracy in terms of the statistical error from the VEGAS Monte Carlo integration is
actually included in the figures. It is better than $1 \%$ though, and thus not
visible on the scale of the plots. The gluon mass regulated graphs were evaluated
over twice as many iterations ($100$) compared to the graphs from Fig. \ref{fig:mcomp}
as the required cancellations are numerically more unstable. In both cases the 
number of evaluations per iteration is $10^6$.

Fig. \ref{fig:caqcfq} displays the sums of all non-Abelian as well as the sum of
all Abelian fermionic contributions to the heavy quark potential. 
In addition we included the one loop corrections (bottom) in the MS-scheme (omitting 
coupling constants) as given in Eq. \ref{eq:pires}. The simple logarithmic
behavior of the massless one loop result is clearly visible and asymptotically
approached by the massive curves. The sign of the one loop correction is opposite to
the two loop Abelian result, reflecting the fact that effectively 
for large momenta $\beta_0^{QED}
\longrightarrow \left(\beta^{QED}_0 \right)^2$ (in the massless case, with 
$\beta_0^{QED}=-\frac{2}{3}$). The relative size of the mass effects are comparable
for the one and two loop corrections.

The massless 
two loop results can be seen
to possess the expected double logarithmic contributions.
The massive two loop results show an almost completely opposite behavior for low 
values of $\frac{m^2}{-q^2}$. At the flavor thresholds, though, both contributions
increase the value obtained from the massless case by the same (relative) 
order of magnitude.
The overall corrections are much larger in absolute terms for the non-Abelian case,
partially due to an extra factor of $C_A$, while in relative terms the Abelian
corrections are bigger. In the high energy regime both graphs show that
the massless limit is approached asymptotically.

The complete massive fermionic two loop contributions 
to the heavy quark potential are
presented in Fig. \ref{fig:cacf}. It can be seen that the overall curve is 
dominated by the non-Abelian threshold behavior (partially due to the extra factor
of $C_A$). The ``$m_c$-graph'' (triangles) matches the massless case for lower
values of $-q^2$ as $m_c^2 \ll m_b^2$. At the respective thresholds we find 
roughly a 33 \% deviation relative to the massless case. This could be very
significant for applications where quark masses are expected to play
an important part.
Furthermore, the physically defined effective charge $\alpha_V({\bf q^2},m^2)$
incorporates quark masses naturally at the flavor thresholds and is also
analytic. Thus, there is no problem of evolving the strong coupling constant
through these thresholds and one never needs to impose matching conditions.
At high values of ${\bf q^2}$ the theory becomes massless and reproduces
the leading logarithmic terms obtained by the $\beta$-function analysis as
these coefficients are scheme independent through two loops in a massless
theory. 

The above analysis can also be helpful for the incorporation of
massive fermions in lattice analyses as the heavy quark potential is defined
by the gauge invariant vacuum expectation value of the Wilson loop in Eq. \ref{eq:Vdef}.
For a direct application of the presented results, a recently proposed way of incorporating
quark flavor thresholds by relating the ``natural'' heavy quark potential 
$m_q$-dependence to an effective continuous and smooth function $n_F(-q^2,m^2)$
\cite{Melles98} seems to be a promising candidate.

\section{Conclusions} \label{sec:con}

We have calculated all the necessary integrals for the non-Abelian massive 
fermionic corrections to the heavy quark potential through two loops.
They describe the
analytic flavor thresholds of the physical coupling $\alpha_V ( {\bf q^2}, m^2 )$.
The new results were obtained by using a mixed analytical, computer-algebraic
as well as numerical
approach and strong consistency checks were performed by observing that all
non-local divergences cancel by adding the appropriate counterterms. 
In case of the complicated two point diagrams
it is found that the weighted sum of all diagrams gives the correct local
gluon wave function renormalization constant. The renormalization constants
were given explicitly. 

It was also checked that no spurious infra-red divergences were introduced by
the implemented reduction scheme as they are present in the intermediate steps
of the calculation. For the explicitly IR-divergent diagrams we proved that
no physical results depend on the introduction of the gluon mass regulator
$\lambda$. This is a consequence of the color singlet state of the external
heavy quark sources.

All physically interesting and gauge invariant finite parts were integrated
with VEGAS \cite{Lepage} and found to agree perfectly with the massless results
of Ref. \cite{Peter96} in that limit which actually checks this part of the 
analysis in \cite{Peter97}. The difference to the massless case
around the charm and bottom flavor thresholds was found to be roughly $33 \%$.
The size of this effect can have important consequences for processes in which
one cannot neglect these masses as well as for the evolution of the strong
coupling constant through analytic flavor thresholds.

\appendix

\section{Decomposition of Two Loop Tensor Integrals} \label{sec:decomp}

For the gluon self energy graphs ${\cal M}_{gse_1}$, ${\cal M}_{gse_4}$ 
and ${\cal M}_{gse_5}$ we chose to not do the fermion
loop integral first, as we did for all vacuum polarization insertions,
but to decompose the occurring tensor integrals into a linear combination
of scalar two loop integrals. The scalar integrals entering in the
expression given in Eq. \ref{eq:pi1t} (or \ref{eq:pi1l}) will be treated in detail in the next
section together with all other integrals needed in this work.

We work in $n$ space-time dimensions, $n=4-\epsilon$, and for the two
loop integrals we use the following notation: 

\begin{equation}
[1] \equiv (l+q)^2-m^2, [2] \equiv l^2-m^2, [3] \equiv (l-k)^2-m^2,
[4] \equiv (k+q)^2, [5] \equiv k^2 \label{eq:denomdef}
\end{equation}

$l$ denotes the loop momentum of
the massive fermion loop and $k$ the remaining loop 
momentum. A prime like $[2'] \equiv l^2$ denotes the massless propagator with
the same momentum as the unprimed. 
While there are different possible technical approaches to our desired decomposition, such
as the one recently suggested in Ref. \cite{Tarasov97},
the general method we use follows that of Ref. \cite{Weiglein94}. We also denote
integrals with squares of ``denominator" terms in the numerator ``$Y$"-
integrals and pure two loop scalar integrals by ``$T$". 

In the following we use various symmetries between $Y$- as well as 
between $T$-integrals. For instance

\begin{equation}
Y^2_{1345}=Y^1_{2345} \;\;\;\;\;,\;\;\;\;\; T_{134}=T_{235}
\end{equation}

For two loop scalar integrals that are actually a product of scalar one loop
integrals we use the respective one loop notation of Ref. \cite{passa79}.
All of the
decompositions were programmed in FORM \cite{verma} and lead to the
following relations for ${\cal M}_{gse_1}$:

\begin{eqnarray}
Y^{11}_{2345} &\equiv& \int \frac{d^nk}{(2 \pi)^n} \int \frac{d^nl}{(2 \pi)^n}
\frac{\mu^{2\epsilon}((l+q)^2-m^2)^2}{(l^2-m^2)((l-k)^2-m^2)(k+q)^2k^2} \nonumber \\
&=& q^2 \left( T_{235}-T_{135} \right) + \frac{1}{n-1} \left[ \frac{n}{6} \left(
-m^4 T_{12'35} +m^2T_{135}+m^2 A_2 B_{12'} - m^2 T_{2'35} - A_2^2\right) 
-m^2 T_{135} \right.
\nonumber \\
&& +m^2 T_{235} + \frac{n}{4} q^4 T_{2345} + \frac{n}{2} q^4 A_2 C_{455}
-m^2 q^4 T_{23455} - q^4 A_2C_{455} + \frac{n}{6} m^2 q^2 T_{12'35} + \frac{7n}{12}
q^2 T_{135} \nonumber \\
&& \left. - \frac{n}{4} q^2 T_{135} + n q^2 A_2B_{45} - \frac{n}{6} q^2 A_2 B_{12'}
+2m^2 q^2 T_{2345} +m^2 q^2 T_{2355} - q^2 T_{135} \right] \label{eq:Y11dec}
\end{eqnarray}

\begin{eqnarray}
Y^{1}_{2345} &\equiv& \int \frac{d^nk}{(2 \pi)^n} \int \frac{d^nl}{(2 \pi)^n}
\frac{\mu^{2\epsilon}((l+q)^2-m^2)}{(l^2-m^2)((l-k)^2-m^2)(k+q)^2k^2} \nonumber \\
&=& A_2 B_{45} + \frac{1}{2} \left( T_{235}-T_{135} + q^2 T_{2345} \right) \\
Y^4_{1235} &\equiv& \int \frac{d^nk}{(2 \pi)^n} \int \frac{d^nl}{(2 \pi)^n}
\frac{\mu^{2\epsilon}(k+q)^2}{((l+q)^2-m^2)(l^2-m^2)((l-k)^2-m^2)k^2} 
\nonumber \\
&=& A_2 B_{12} + q^2 T_{1235} + \frac{1}{2} \left( T_{2'35} + A_2 B_{22'} - A_2 B_{12'}
-T_{135}-q^2 A_2 C_{122'}+(m^2-q^2) T_{12'35} \right) \\
Y^{3}_{1245} &\equiv& \int \frac{d^nk}{(2 \pi)^n} \int \frac{d^nl}{(2 \pi)^n}
\frac{\mu^{2\epsilon}((l-k)^2-m^2)}{((l+q)^2-m^2)(l^2-m^2)(k+q)^2k^2} 
\;=\; A_2B_{45}- \frac{q^2}{2} B_{12}B_{45} \\
Y^{1}_{234} &\equiv& \int \frac{d^nk}{(2 \pi)^n} \int \frac{d^nl}{(2 \pi)^n}
\frac{\mu^{2\epsilon}((l+q)^2-m^2)}{(l^2-m^2)((l-k)^2-m^2)(k+q)^2} \nonumber \\
&=& \frac{1}{3} \left[ A_2^2 + m^2 \left( T_{2'35}-A_2B_{12'}-T_{135}+m^2 
T_{12'35} \right) + q^2 \left(T_{135}+A_2B_{12'}-m^2 T_{12'35} \right) \right] \\
Y^{4}_{135} &\equiv& \int \frac{d^nk}{(2 \pi)^n} \int \frac{d^nl}{(2 \pi)^n}
\frac{\mu^{2\epsilon}(k+q)^2}{((l+q)^2-m^2)((l-k)^2-m^2)k^2} \nonumber \\
&=& \frac{1}{3} \left( A_2^2 + q^2 T_{135} \right) - \frac{2}{3} \left[ 
m^2 \left( T_{2'35}-A_2B_{12'}-T_{135}+m^2 
T_{12'35} \right) + q^2 \left(A_2B_{12'}-m^2 T_{12'35} \right) \right] \\
Y^{1}_{235} &\equiv& \int \frac{d^nk}{(2 \pi)^n} \int \frac{d^nl}{(2 \pi)^n}
\frac{\mu^{2\epsilon}((l+q)^2-m^2)}{(l^2-m^2)((l-k)^2-m^2)k^2} 
\;=\; q^2 T_{235} \\
Y^{1}_{245} &\equiv& \int \frac{d^nk}{(2 \pi)^n} \int \frac{d^nl}{(2 \pi)^n}
\frac{\mu^{2\epsilon}((l+q)^2-m^2)}{(l^2-m^2)(k+q)^2k^2} 
\;=\; q^2 A_2B_{45} 
\end{eqnarray}

For the remaining two diagrams, ${\cal M}_{gse_4}$ and ${\cal M}_{gse_5}$,
we have slightly different denominators as is evident from 
Eqs. \ref{eq:gse4def} and \ref{eq:gse5def}. It is possible, though, to
relate these to the conventions given in \ref{eq:denomdef} with the
exception of the finite scalar integral $T^A_{12345}$ which is given in
Eq. \ref{eq:TA12345}. ``$A$'' denotes the fact that the topology of these
diagrams is Abelian. Below we list the $Y$-functions we need for the
required decomposition with terms on the l.h.s. having the denominators
of the original integrals and given in terms of functions on the r.h.s.
which are using the conventions of Eq. \ref{eq:denomdef}:

\begin{eqnarray}
\mbox{$^AY_{2345}^1$} &=& A_2 B_{12}+T_{235}-T_{135}+q^2 T_{1235}
-Y^4_{1235} \\
\mbox{$^AY_{1245}^3$} &=& 2 A_2 B_{12} + (2m^2-\frac{q^2}{2}) B^2_{12}
\\ \mbox{$^AY_{2335}^4$} &=& T_{235} +q^2 T_{2235} 
\\ \mbox{$^AY_{135}^2$} &=& \mbox{$^AY_{234}^5$}=\mbox{$^AY_{135}^4$}=
\mbox{$^AY_{234}^1$}=Y^1_{234} \\
\mbox{$^AY_{235}^4$} &=& q^2 T_{235} \\
\mbox{$^AY_{235}^1$} &=& q^2 T_{235} \\
\mbox{$^AY_{255}^4$} &=& A_2^2 + q^2 A_2 B_{22} 
\end{eqnarray}

\section{Two-Loop Integrals} \label{sec:tli}

In this appendix we give the explicit results for all the integrals needed in
the calculation of the two loop fermionic corrections to the heavy quark 
potential. These include all the scalar two loop integrals occurring in the
decomposition of the gluon self energy graph ${\cal M}_{gse_1}$ in section \ref{sec:res} as
well as the remaining tensor integrals needed for the remaining contributions.
Since the potential between two infinitely heavy color test charges represents
a physical quantity, all integrals presented are real due to the spacelike
value of the physical momentum transfer $q^2 < 0$. For this reason we found
it convenient to adopt both analytical as well as numerical methods for
the implementation into FORTRAN. Wherever possible we proceed with the 
integration of the remaining Feynman parameter integrals and where this 
becomes too involved, we integrate the remainder with the Monte Carlo
integrator VEGAS, \cite{Lepage}.

The notation is as follows:

The following Feynman parameter identities \cite{PascTar} are very useful and were employed
in all integrals in this work:

\begin{eqnarray}
\frac{1}{a_1...a_m} &=& \Gamma(m) \!\! \int^1_0 \!\!\!\!du_1 ... \!\!
\int^1_0 \!\!\!\!du_{m-1} 
\frac{u_1^{m-2} u_2^{m-3}... \; u_{m-2}}{(a_1u_1...u_{m-1}+a_2u_1...u_{m-2}(1-u_{m-1})
+ ...+ a_m(1-u_1))^m} \label{eq:Feynp1} \\
\frac{1}{a^\alpha b^\beta} &=& \frac{\Gamma(\alpha+\beta)}{\Gamma(\alpha)\Gamma(\beta
)} \int^1_0 \!\!\!\!du \frac{u^{\alpha-1}(1-u)^{\beta-1}}{(a u +b (1-u))^{\alpha+\beta}}
\label{eq:Feynp2} \\
\frac{1}{a^\alpha b^\beta c^\gamma} &=& \frac{\Gamma(\alpha+\beta +\gamma)}
{\Gamma(\alpha)\Gamma(\beta)\Gamma(\gamma
)} \int^1_0 \!\!\!\!du \;u \int^1_0 \!\!\!\!dv \frac{(uv)^{\alpha-1}(u(1-v))^{\beta-1}
(1-u)^{\gamma-1}}{(a u v +b u(1-v) + c (1-u))^{\alpha+\beta +\gamma}}
\label{eq:Feynp3} \\ 
\frac{1}{a^\alpha b^\beta c^\gamma d^\delta} &=& \frac{\Gamma(\alpha+\beta +\gamma +\delta)}
{\Gamma(\alpha)\Gamma(\beta)\Gamma(\gamma) \Gamma(\delta
)} \! \int^1_0 \!\!\!\!du \;u^2 \!\!\! \int^1_0 \!\!\!\!dv \;v \!\! \int^1_0 \!\!\!\!dw
\frac{(uvw)^{\alpha-1}(uv(1-w))^{\beta-1}
(u(1-v))^{\gamma-1}(1-u)^{\delta-1}}{(a u v w +b uv(1-w) + c u (1-v) 
+ d (1-u))^{\alpha+\beta +\gamma +\delta}}
\label{eq:Feynp4} 
\end{eqnarray}

We use the following abbreviations in addition:

\begin{eqnarray}
\eta &\equiv& \frac{4 \pi \mu^2}{m^2} , \;\; \zeta \equiv \frac{4 \pi \mu^2}{-q^2} , \;\;
\alpha \equiv u+(1-u)x(1-x) , \;\; \widetilde{\alpha} \equiv u+(1-u)(1-x) 
\label{eq:paramdef} \\
\Delta &\equiv& \frac{q^2}{m^2} \left( \frac{u^2(1-v)^2}{\alpha^2}-
\frac{u(1-v)}{\alpha} \right) + \frac{1-u}{\alpha} \label{eq:Delta} \\
\widetilde{\Delta} &\equiv& \frac{q^2}{m^2} \left( \frac{u^2(1-v)^2}{
\widetilde{\alpha}^2}-
\frac{u(1-v)}{\widetilde{\alpha}} \right) + \frac{1}{\widetilde{\alpha}} 
\label{eq:Deltap} \\
\widetilde{\Delta'} &\equiv& \frac{q^2}{m^2} \left( \frac{u^2(1-v)^2}{
\widetilde{\alpha}^2}-
\frac{u(1-v)}{\widetilde{\alpha}} \right) + \frac{1-u v}{\widetilde{\alpha}} 
\label{eq:Deltapp}
\end{eqnarray}

where $\mu$ is the dimensional-regularization mass parameter \cite{PeskSchr}.
All results are given in terms of their
dependence on $\epsilon$ and would have to be expanded with the factors given
in the explicit results of section \ref{sec:res} up to ${\cal O} \left( \epsilon
\right)$. The results in this paper were obtained by employing MAPLE to do the
required expansion and are too cumbersome for explicit presentation.

We start with results of the following simple scalar one and two loop 
functions:

\begin{eqnarray}
A_2 &\equiv&  \int \frac{d^nl}{(2 \pi)^n} \frac{\mu^{\epsilon}}{(l^2-m^2)} =
- \frac{i\; m^2\eta^\frac{\epsilon}{2} \Gamma \left( -1 + \frac{\epsilon}{2} \right)
}{16 \pi^2} \label{eq:A2}\\ 
B_{22} &\equiv&  \int \frac{d^nl}{(2 \pi)^n} \frac{\mu^{\epsilon}}{(l^2-m^2)^2} =
\frac{i\; \eta^\frac{\epsilon}{2} \Gamma \left( \frac{\epsilon}{2} \right)
}{16 \pi^2} \label{eq:B22} \\ 
B_{22'} &\equiv&  \int \frac{d^nl}{(2 \pi)^n} \frac{\mu^{\epsilon}}{(l^2-m^2)l^2} =
\frac{i\; \eta^\frac{\epsilon}{2} \Gamma \left( \frac{\epsilon}{2} \right)
}{16 \pi^2 \left( 1- \frac{\epsilon}{2} \right)} \label{eq:B22m} \\ 
B_{12'} &\equiv&  \int \frac{d^nl}{(2 \pi)^n} \frac{\mu^{\epsilon}}{((l+q)^2-
m^2)l^2} \;=\; \int^1_0 dx \frac{i \eta^{\frac{\epsilon}{2}} \Gamma \left( \frac{ \epsilon}
{2} \right)}{(4 \pi)^2 (\frac{-q^2}{m^2}x (1-x)+x)^\frac{\epsilon}{2}} \label{eq:B12m} \\
B_{12} &\equiv&  \int \frac{d^nl}{(2 \pi)^n} \frac{\mu^{\epsilon}}{((l+q)^2-
m^2)(l^2-m^2)} \;=\; \int^1_0 dx \frac{i \eta^{\frac{\epsilon}{2}} \Gamma \left( \frac{ 
\epsilon}{2}
\right)}{(4 \pi)^2 (\frac{-q^2}{m^2}x (1-x)+1)^\frac{\epsilon}{2}} \label{eq:B12} \\
B_{45} &\equiv&  \int \frac{d^nk}{(2 \pi)^n} \frac{\mu^{\epsilon}}{(k+q)^2
k^2} \;=\; \int^1_0 dx \frac{i \eta^{\frac{\epsilon}{2}} \Gamma \left( \frac{ \epsilon}
{2} \right)}{(4 \pi)^2 (\frac{-q^2}{m^2}x (1-x))^\frac{\epsilon}{2}} 
= \frac{ i \zeta^\frac{\epsilon}{2} \Gamma \left( \frac{\epsilon}{2} \right)
\Gamma^2 \left( 1- \frac{\epsilon}{2} \right)}
{(4 \pi)^2 \Gamma \left( 2- \epsilon \right)} \label{eq:B45} \\
C_{455} &\equiv&  \int \frac{d^nk}{(2 \pi)^n} \frac{\mu^{\epsilon}}{(k+q)^2
k^4} 
= \frac{ i \zeta^\frac{\epsilon}{2} \Gamma \left( - \frac{\epsilon}{2} \right)
\Gamma \left( 1-\frac{\epsilon}{2} \right)
\Gamma \left( 1+ \frac{\epsilon}{2} \right)}
{q^2 (4 \pi)^2 \Gamma \left( 1- \epsilon \right)} \label{eq:C455} \\
C_{122} &\equiv&  \int \frac{d^nl}{(2 \pi)^n} \frac{\mu^{\epsilon}}{((l+q)^2-
m^2)(l^2-m^2)^2} \;=\; -\int^1_0 dx \frac{i \;x \eta^{\frac{\epsilon}{2}} \Gamma 
\left(1+ \frac{ \epsilon}{2}
\right)}{(4 \pi)^2 m^2 (\frac{-q^2}{m^2}x (1-x)+1)^{1+\frac{\epsilon}{2}}} \label{eq:C122} \\
C_{122'} &\equiv&  \int \frac{d^nl}{(2 \pi)^n} \frac{\mu^{\epsilon}}{((l+q)^2-
m^2)(l^2-m^2)l^2} \;=\; - \!\! \int^1_0 \!\!dx \int^1_0 \!\!dy \frac{i \eta^{\frac{\epsilon}{2}} \Gamma 
\left(1+ \frac{ \epsilon}{2}
\right) x^{-\frac{\epsilon}{2}}}{(4 \pi)^2 m^2 (\frac{q^2}{m^2}(x (1-y)^2-1+y)+1)^{1+
\frac{\epsilon}{2}}} \label{eq:C122'} 
\end{eqnarray}

A very useful integral for \ref{eq:C122'} is given by

\begin{equation}
\int^1_0 dx \int^1_0 dy \frac{1}{(a(x (1-y)^2-1+y)+1)} = - \frac{2}{a} \sqrt{a^2-4 a} \;\;
tanh^{-1} \left( \sqrt{\frac{a}{a-4}} \right) - \frac{a-1}{a} log (1-a)
\end{equation}

This integral is needed in order to analytically separate the divergent pieces since
$C_{122'}$ is multiplied by $A_2$ in the solution for Eq. \ref{eq:gse4def}.
 
\begin{eqnarray}
T_{2'35} &\equiv& \int \frac{d^nk}{(2 \pi)^n} \int \frac{d^nl}{(2 \pi)^n}
\frac{\mu^{2\epsilon}}{l^2 ((l-k)^2-m^2)k^2} 
= \frac{m^2 \eta^{\epsilon} \Gamma \left( \frac{\epsilon}{2} \right)
\Gamma \left( -1+\epsilon \right) \Gamma^2 \left( 1- \frac{\epsilon}{2} \right)}
{(4 \pi)^4 \Gamma \left( 2- \frac{\epsilon}{2} \right)} \label{eq:T2'35} \\
T_{235} &\equiv& \int \frac{d^nk}{(2 \pi)^n} \int \frac{d^nl}{(2 \pi)^n}
\frac{\mu^{2\epsilon}}{(l^2-m^2)((l-k)^2-m^2)k^2} 
= \frac{m^2 \eta^{\epsilon} \Gamma^2 \left( \frac{\epsilon}{2} \right)
\Gamma \left( -1+\epsilon \right) \Gamma \left( 1- \frac{\epsilon}{2} \right)}
{(4 \pi)^4 \Gamma \left( 2- \frac{\epsilon}{2} \right)\Gamma \left(
\epsilon \right)} \label{eq:T235}
\end{eqnarray}

The reason why the following integrals cannot be given in such a simple form
is the presence of the external momentum transfer $q$ in addition to the
masses. In order to extract the infinite pieces from the next integral $T_{135}$,
we repeatedly use the following propagator identity:

\begin{equation}
\frac{1}{(l+q)^2-m^2} \; = \; \frac{1}{l^2-m^2} \; - \; \frac{2lq+q^2}{(l^2-m^2)((l+q)^2-m^2)} \label{eq:propid}
\end{equation}
 
It then follows that

\begin{equation}
T_{135} \equiv \int \frac{d^nk}{(2 \pi)^n} \int \frac{d^nl}{(2 \pi)^n}
\frac{\mu^{2\epsilon}}{((l+q)^2-m^2)((l-k)^2-m^2)k^2} = T_{235} +T_a +T_b + T_c \;\;\;,
\end{equation}
 
with

\begin{equation}
T_a \equiv - \int \frac{d^nk}{(2 \pi)^n} \int \frac{d^nl}{(2 \pi)^n}
\frac{\mu^{2\epsilon}(2lq+q^2)}{(l^2-m^2)^2((l-k)^2-m^2)k^2} 
= \frac{q^2 \eta^{\epsilon} \Gamma \left( \frac{\epsilon}{2} \right)
\Gamma \left( \epsilon \right) \Gamma \left( 1- \frac{\epsilon}{2} \right)
\Gamma \left( 1+\frac{\epsilon}{2} \right)}
{(4 \pi)^4 \Gamma \left( 2- \frac{\epsilon}{2} \right)\Gamma \left( 1 +
\epsilon \right)} \label{eq:Ta}
\end{equation}

\begin{eqnarray}
T_b &\equiv&  \int \frac{d^nk}{(2 \pi)^n} \int \frac{d^nl}{(2 \pi)^n}
\frac{\mu^{2\epsilon}(2lq+q^2)^2}{(l^2-m^2)^3((l-k)^2-m^2)k^2} 
= -\frac{4q^2 \eta^{\epsilon} \Gamma \left( \frac{\epsilon}{2} \right)
\Gamma \left( \epsilon \right) \Gamma \left( 1- \frac{\epsilon}{2} \right)
\Gamma \left( 1+\frac{\epsilon}{2} \right)}
{n (4 \pi)^4 \Gamma \left( 2- \frac{\epsilon}{2} \right)\Gamma \left( 1 +
\epsilon \right)} \nonumber \\
&&+\frac{q^2\left( \frac{4}{n}+\frac{q^2}{m^2}\right) \eta^{\epsilon} 
\Gamma \left( \frac{\epsilon}{2} \right)
\Gamma \left(1+ \epsilon \right) \Gamma \left( 1- \frac{\epsilon}{2} \right)
\Gamma \left( 2+\frac{\epsilon}{2} \right)}
{2 (4 \pi)^4 \Gamma \left( 2- \frac{\epsilon}{2} \right)\Gamma \left( 2 +
\epsilon \right)} \label{eq:Tb}
\end{eqnarray}

In passing we note that 

\begin{equation}
T_{2235} = - \frac{1}{q^2} T_a
\end{equation}

The last term $T_c$ has only a simple pole in $\epsilon$ which is, however,
buried in the Feynman parameter integration. This is a quite common problem
that arises because of the $\Gamma$-factors in Eqs. \ref{eq:Feynp2} and
\ref{eq:Feynp3}. We take ``$u$" to be that Feynman parameter and for our 
purposes it suffices to write the following identity:

\begin{equation}
\int^1_0 du (1-u)^{\frac{\epsilon}{2}-1} f(u) \;=\; \frac{2}{\epsilon} f(1) + 
\int^1_0 du (1-u)^{\frac{\epsilon}{2}-1} \left( f(u) - f(1) \right) \label{eq:fuid}
\end{equation} 

The respective terms for $T_c \equiv -\int \frac{d^nk}{(2 \pi)^n} \int \frac{d^nl}
{(2 \pi)^n}\frac{\mu^{2\epsilon}(2lq+q^2)^3}{(l^2-m^2)^3((l+q)^2-m^2)
((l-k)^2-m^2)k^2}$ are:

\begin{eqnarray}
f(u) &\equiv& \int^1_0 dx \int^1_0 dv \frac{q^2 \eta^\epsilon u^3 v^2}{2 (4 \pi)^4 x^{\frac{\epsilon}{2}}
\widetilde{\alpha}^{4+\frac{\epsilon}{2}}} \left( \frac{q^4}
{m^4} \frac{1-8 \frac{u^3 (1-v)^3}{\widetilde{\alpha}^3} +12 \frac{u^2 (1-v)^2
}{\widetilde{\alpha}^2} - 6 \frac{u (1-v)}{\widetilde{\alpha}}}{\widetilde{
\Delta}^{2+\epsilon}} \Gamma(2+\epsilon) \right. \nonumber \\
&& - \left. \frac{q^2}{m^2} \frac{6-12 \frac{u (1-v)}{
\widetilde{\alpha}}}{\widetilde{\Delta}^{1+\epsilon}} \Gamma(1+\epsilon)
\right) \label{eq:Tcfu}
\end{eqnarray}

and thus

\begin{eqnarray}
f(1) &\equiv& \int^1_0 dx \int^1_0 dv \frac{q^2 \eta^\epsilon v^2}{2 (4 \pi)^4 x^{\frac{\epsilon}{2}}
} \left( \frac{q^4}
{m^4} \frac{1-8 (1-v)^3 +12 (1-v)^2
 - 6 (1-v)}{\left(- \frac{q^2}{m^2} v (1-v) +1 \right)^{2+\epsilon}}
\Gamma(2+\epsilon) \right. \nonumber \\
&&- \left. \frac{q^2}{m^2} \frac{6-12 (1-v)
}{\left( - \frac{q^2}{m^2} v (1-v) +1 \right)^{1+\epsilon}} \Gamma(1+\epsilon)
\right) \label{eq:Tcf1}
\end{eqnarray}

Although this result for $T_{135}$ is correct, it is numerically 
unstable in the massless limit because of terms of order 
$\frac{q^2}{m^2}$ which have to cancel as $m^2 \longrightarrow 0$. 
A way out of this calamity as well as a very good check on the 
correctness of our result for this integral is to use the propagator
identity \ref{eq:propid} for $\frac{1}{(k+q)^2}$ instead after shifting
the loop momenta. This yields

\begin{equation}
T_{135} = T_{235} - q^2 T_{2345} - <<\frac{2kq}{[2][3][4][5]}>>
\end{equation}

The result for $T_{2345}$ is given below and the last term in the
equation can easily be found to be $2 q^2 \frac{u (1-v)}{\alpha}$ times
the expressions for the scalar integral. This term just stems from
the momentum shift $k \longrightarrow k'- q \frac{u(1-v)}{\alpha}$.
Numerically, away from the singularity at $m=0$, both solutions agree.

In similar ways we treat the following more complicated integrals, always 
calling ``$u$" the Feynman parameter that contains an additional divergence
if $f(u)$-terms are quoted. The desired value for the respective integrals
are understood to follow from an expansion in $\epsilon$ of Eq. \ref{eq:fuid}.
For

\begin{equation}
T_{2345} \equiv \int \frac{d^nk}{(2 \pi)^n} \int \frac{d^nl}{(2 \pi)^n}
\frac{\mu^{2\epsilon}}{(l^2-m^2)((l-k)^2-m^2)(k+q)^2k^2} \label{eq:T2345def}
\end{equation}

we get

\begin{equation}
f(u) \equiv -\int^1_0 dx \int^1_0 dv \frac{\eta^\epsilon \Gamma(\epsilon) u}{(4 \pi)^4 \alpha^{2+\frac
{\epsilon}{2}} \Delta^\epsilon}\; , \;\;
f(1) \equiv - \int^1_0 dv \frac{\eta^\epsilon \Gamma(\epsilon)}{(4 \pi)^4 
(-\frac{q^2}{m^2} v (1-v))^\epsilon}  \label{eq:T2345fu}
\end{equation}

Similarly,

\begin{equation}
T_{1235} \equiv \int \frac{d^nk}{(2 \pi)^n} \int \frac{d^nl}{(2 \pi)^n}
\frac{\mu^{2\epsilon}}{((l+q)^2-m^2)(l^2-m^2)((l-k)^2-m^2)k^2} \label{eq:T1235def}
\end{equation}

with

\begin{equation}
f(u) \equiv -\int^1_0 dx \int^1_0 dv \frac{\eta^\epsilon \Gamma(\epsilon) u}{(4 \pi)^4 x^{\frac{\epsilon
}{2}} \widetilde{\alpha}^{2+\frac
{\epsilon}{2}} \widetilde{\Delta}^\epsilon}\; , \;\;
f(1) \equiv -\int^1_0 dx \int^1_0 dv \frac{\eta^\epsilon \Gamma(\epsilon)}{(4 \pi)^4 x^\frac{\epsilon}{2}
(-\frac{q^2}{m^2} v (1-v)+1)^\epsilon}  \label{eq:T1235fu}
\end{equation}

For

\begin{equation}
T_{12'35} \equiv \int \frac{d^nk}{(2 \pi)^n} \int \frac{d^nl}{(2 \pi)^n}
\frac{\mu^{2\epsilon}}{((l+q)^2-m^2)l^2((l-k)^2-m^2)k^2} \label{eq:T12m35def}
\end{equation}

we find

\begin{equation}
f(u) \equiv -\int^1_0 dx \int^1_0 dv \frac{\eta^\epsilon \Gamma(\epsilon) u}{(4 \pi)^4 x^{\frac{\epsilon
}{2}} \widetilde{\alpha}^{2+\frac
{\epsilon}{2}} \widetilde{\Delta'}^\epsilon}\; , \;\;
f(1) \equiv - \int^1_0 dx \int^1_0 dv \frac{\eta^\epsilon \Gamma(\epsilon)}{(4 \pi)^4 x^\frac{\epsilon}{2}
(-\frac{q^2}{m^2} v (1-v)+1-v)^\epsilon}  \label{eq:T12m35fu}
\end{equation}

The infra-red finite integral 

\begin{equation}
I_{2455} \equiv - \int \frac{d^nk}{(2 \pi)^n} \int \frac{d^nl}{(2 \pi)^n}
\frac{\mu^{2\epsilon}(k^2+2kq)}{(l^2-m^2)(k+q)^2k^4} \label{eq:I2455}
\end{equation}

is a product of two one loop functions which are given by

\begin{eqnarray}
A_2 &\equiv& - \frac{i m^2 \eta^\frac{\epsilon}{2} \Gamma \left(-1+
\frac{\epsilon}{2} \right)}
{(4 \pi)^2} \\
I_{455} &\equiv& - \int^1_0 du  \frac{i \zeta^\frac{\epsilon}{2}}{(4 \pi)^2} \left( \frac{
n}{2(u(1-u))^\frac{\epsilon}{2}} \Gamma \left(\frac{\epsilon}{2} \right)
- \frac{(1-u)(1+u)}{(u(1-u))^{1+\frac{\epsilon}{2}}} \Gamma \left( 1 +
\frac{\epsilon}{2} \right) \right) \label{eq:I455}
\end{eqnarray}

and in dimensional regularization we have $ I_{2455}= A_2 I_{455}=q^2 A_2 C_{455}$, where
$A \& C$ denote the respective one loop scalar integrals.
For the infra-red finite combination 

\begin{equation}
I_{23455} \equiv q^2 T_{23455}-T_{2355} \;=\; -\int \frac{d^nk}{(2 \pi)^n} \int 
\frac{d^nl}{(2 \pi)^n} \frac{\mu^{2\epsilon}(k^2+2kq)}{(l^2-m^2)((l-k)^2-m^2)(k+q)^2k^4}
\label{eq:I23455}
\end{equation}

we get two ``$f(u)$" terms, distinguished below by capital (containing double
pole terms) and lower case (with only simple poles) letters:

\begin{eqnarray}
F(u) &\equiv& \int^1_0 dx \int^1_0 dv \frac{n \eta^\epsilon \Gamma(\epsilon) u^2 v}{2 (4 \pi)^4 \alpha^{3+\frac
{\epsilon}{2}} \Delta^\epsilon}\; , \;\;
F(1) \equiv \int^1_0 dv \frac{n \eta^\epsilon \Gamma(\epsilon) v}{2(4 \pi)^4 
(-\frac{q^2}{m^2} v (1-v))^\epsilon}  \label{eq:I23455Fu} \\
f(u) &\equiv& \int^1_0 dx \int^1_0 dv A \frac{ \eta^\epsilon \Gamma(1+\epsilon) u^2 v}{(4 \pi)^4 \alpha^{3+\frac
{\epsilon}{2}} \Delta^{1+\epsilon}}\; , \;\;
f(1) \equiv - \int^1_0 dv \frac{(1+v) \eta^\epsilon \Gamma(1+\epsilon)}{(4 \pi)^4 
(-\frac{q^2}{m^2} v (1-v))^\epsilon}  \label{eq:I23455fu} \\
A &\equiv& -\frac{q^2}{m^2} \left( \frac{u^2(1-v)^2}{\alpha^2} -2 \frac{u(1-v)}
{\alpha} \right) \label{eq:I23455Ff}
\end{eqnarray}

For

\begin{equation}
T_{12235} \equiv \int \frac{d^nk}{(2 \pi)^n} \int \frac{d^nl}{(2 \pi)^n}
\frac{\mu^{2\epsilon}}{((l+q)^2-m^2)(l^2-m^2)^2((l-k)^2-m^2)k^2} \label{eq:T12235def}
\end{equation}

we find

\begin{equation}
f(u) \equiv \int^1_0 dx \int^1_0 dv \frac{\eta^\epsilon \Gamma(1+\epsilon) u^2v}{(4 \pi)^4 x^{\frac{\epsilon
}{2}} \widetilde{\alpha}^{3+\frac
{\epsilon}{2}} \widetilde{\Delta}^{1+\epsilon}}\; , \;\;
f(1) \equiv \int^1_0 dx \int^1_0 dv \frac{\eta^\epsilon \Gamma(1+\epsilon) v}{(4 \pi)^4 x^\frac{\epsilon}{2}
(-\frac{q^2}{m^2} v (1-v)+1)^{1+\epsilon}}  \label{eq:T12235fu}
\end{equation}

The completely finite integral $T_{12345} \equiv \int \frac{d^nk}{(2 \pi)^n} \int
\frac{d^nl}{(2 \pi)^n} \frac{\mu^{2\epsilon}}{(l^2-m^2)((l+q)^2-m^2)((l-k)^2-m^2)(k+q)^2k^2} $ is given by:

\begin{equation}
  T_{12345} \;=\; \int^1_0 dx \int^1_0 dy \int^1_0 du \int^1_0 dv \frac{\eta^\epsilon \Gamma(1+\epsilon)}{m^2 (4 \pi)^4} \frac{
x u (1-u)^\frac{\epsilon}{2}}{\alpha^{3+ \frac{\epsilon}{2}} \left( \frac{q^2}{m^2} 
\left( \frac{\sigma^2}{\alpha^2}- \frac{\rho}{\alpha} \right) + \frac{
\widetilde{\alpha}}{\alpha} \right)^{1+ \epsilon}} \label{eq:T12345}
\end{equation}

$\epsilon$ can of course be set to zero in the above expression and we use
the following abbreviations:

\begin{eqnarray}
\sigma &\equiv& u(1-v)+(1-u)(1-y)x(1-x) \label{eq:sigma} \\
\rho &\equiv& u(1-v)+(1-u)(x(1-y)-x^2(1-y)^2) \label{eq:rho}
\end{eqnarray}

For the ``Abelian'' gluon self energy graph ${\cal M}_{gse_4}$ we need another
completely finite integral with five denominators, namely
$T^A_{12345} \equiv \int \frac{d^nk}{(2 \pi)^n} \int
\frac{d^nl}{(2 \pi)^n} \frac{\mu^{2\epsilon}}{(l^2-m^2)((l+q)^2-m^2)(l-k)^2((k+q
)^2-m^2)(k^2-m^2)}$. Here we find
 
\begin{equation}
  T^A_{12345} \;=\; \int^1_0 dx \int^1_0 dy \int^1_0 du \int^1_0 dv \frac{\eta^\epsilon \Gamma(1+\epsilon)}{m^2 (4 \pi)^4} \frac{
x u (1-u)^\frac{\epsilon}{2}}{\alpha^{3+ \frac{\epsilon}{2}} \left( \frac{q^2}{m^2} 
\left( \frac{\sigma^2}{\alpha^2}- \frac{\rho}{\alpha} \right) + \frac{
x (1-u)}{\alpha} \right)^{1+ \epsilon}} \label{eq:TA12345}
\end{equation}

Again, we can savely set $\epsilon$ to zero like above.
The following integrals are needed for the diagrams where we integrated out the
fermion loop first, with
$\pi (k^2,m^2)$ taken from Eq. \ref{eq:pires}:

\begin{eqnarray}
\int \frac{d^nk}{(2 \pi)^n} \frac{\mu^\epsilon \pi (k^2,m^2)}{k^2} &=& \int^1_0 dx \int^1_0 du \frac{m^2 \Gamma 
\left(-1+\epsilon \right) x (1-x) (1-u)^{-\frac{\epsilon}{2}} \eta^\epsilon}
{32 \pi^4 \alpha^{2-\frac{\epsilon}{2}}} \label{eq:I1} \\
\int \frac{d^nk}{(2 \pi)^n} \frac{\mu^\epsilon \pi (k^2,m^2)}{(k+q)^2} &=& \int^1_0 dx \int^1_0 du \frac{m^2 \Gamma
\left(-1+\epsilon \right) x (1-x) (1-u)^{-\frac{\epsilon}{2}}\eta^\epsilon
\left(-\frac{q^2}{m^2} \frac{ux(1-x)}{\alpha}+1 \right)^{1-\epsilon}}
{32 \pi^4 \alpha^{2-\frac{\epsilon}{2}}} \label{eq:I2} \\
\int \frac{d^nk}{(2 \pi)^n} \frac{ 2kq \mu^\epsilon \pi (k^2,m^2)}{(k+q)^2} &=& \int^1_0 dx \int^1_0 du \frac{-q^2 m^2
\Gamma \left(-1+\epsilon \right) x (1-x) u (1-u)^{-\frac{\epsilon}{2}}\eta^\epsilon
\left(-\frac{q^2}{m^2} \frac{ux(1-x)}{\alpha}+1 \right)^{1-\epsilon}}
{16 \pi^4 \alpha^{3-\frac{\epsilon}{2}}} \label{eq:I3} 
\end{eqnarray}

Below we split again into $f(u)$ and $f(1)$ terms. For 

\begin{equation}
\int \frac{d^nk}
{(2 \pi)^n} \frac{\mu^\epsilon \pi (k^2,m^2)}{(k+q)^2k^2} \label{eq:I4}
\end{equation}

we find:

\begin{eqnarray}
f(u) &=& - \int^1_0 dx \int^1_0 dv \frac{
\Gamma \left(\epsilon \right) u x (1-x)\eta^\epsilon }{32 \pi^4  
\alpha^{2+\frac{\epsilon}{2}} \Delta^\epsilon} \label{eq:I4fu} \\ 
f(1) &=& - \int^1_0 dx \int^1_0 dv \frac{
\Gamma \left(\epsilon \right)x (1-x)\eta^\epsilon }{32 \pi^4 \left( \frac{ 
-q^2 v (1-v)}{m^2} \right)^\epsilon} \label{eq:I4f1} 
\end{eqnarray}

For 

\begin{equation}
\int \frac{d^nk}
{(2 \pi)^n} \frac{(k^2+2kq) \mu^\epsilon \pi (k^2,m^2)}{(k+q)^2k^4} \label{eq:I5}
\end{equation}

there are two contributions
corresponding to terms with double poles ($F$) and only single poles ($f$):

\begin{eqnarray}
F(u) &\equiv& - \int^1_0 dx \int^1_0 dv \frac{\Gamma \left(\epsilon \right)n u^2v x (1-x)\eta^\epsilon }
{64 \pi^4 \alpha^{3+\frac{\epsilon}{2}} \Delta^\epsilon} \label{eq:I5Fu} \\
F(1) &=& - \int^1_0 dx \int^1_0 dv \frac{
\Gamma \left(\epsilon \right)n v x (1-x)\eta^\epsilon }{64 \pi^4 \left( \frac{ 
-q^2 v (1-v)}{m^2} \right)^\epsilon} \label{eq:I5F1} \\
f(u) &\equiv& \int^1_0 dx \int^1_0 dv \frac{q^2\Gamma \left(1+ \epsilon \right) u^2v x (1-x)\eta^\epsilon
\left( \frac{u^2(1-v)^2}{\alpha^2}-2\frac{u(1-v)}{\alpha} \right) }
{32 m^2 \pi^4 \alpha^{3+\frac{\epsilon}{2}} \Delta^{1+\epsilon}} \label{eq:I5fu} \\
f(1) &=& \int^1_0 dx \int^1_0 dv \frac{\Gamma \left( 1+ \epsilon \right) (1+v) x (1-x) \eta^\epsilon}
{32 \pi^4 \left( \frac{-q^2v(1-v)}{m^2} \right)^\epsilon} \label{eq:I5f1} 
\end{eqnarray}

\subsection{Two Loop Integrals with Gluon Mass} \label{sec:tlgm}

In this appendix we give details about the evaluation of the IR-divergent
integrals of section \ref{sec:IR}. The contributions containing heavy quark 
propagator terms were regulated using a gluon mass regulator and lead to
the following general integral over $k_0$:

\begin{equation}
I_{k_0} \equiv \int^\infty_{-\infty} \frac{d k_0}{2 \pi} \frac{1}{(k_0 + i \varepsilon)^2
(-k_0^2+ {\bf k}^2 + M^2 - i \varepsilon)^\beta} \label{eq:k0int}
\end{equation}

The general power in integral \ref{eq:k0int} leads to a branch cut along
the real axis for all those values for which $k_0^2 \ge {\bf k}^2 + M^2$.
Including the $i \varepsilon$-prescription as indicated in \ref{eq:k0int}, we
choose a path in the complex plane around the branch cut in the upper half
of the plane and find the following solution:

\begin{eqnarray}
I_{k_0} &=& -2 \; i \; sin \left( \beta
\pi \right) \int^\infty_{\sqrt{{\bf k}^2+M^2}} \frac{d k_0}{2 \pi}
\frac{1}{k_0^2 |-k_0^2+{\bf k}^2+M^2 |^\beta} \nonumber \\
&=& - 2 \; i \; sin \left( \beta \pi \right) \frac{ \Gamma \left( 
1 - \beta \right) \Gamma \left( \frac{1}{2} +\beta \right)}{ 2 \pi^\frac{3}{2}
( {\bf k}^2+M^2 )^{\frac{1}{2} +\beta}} \label{eq:k0intres}
\end{eqnarray}

The remaining Euclidean integral can then be performed easily. In the case
of the crossed ladder diagram ${\cal M}_{cl}$ we find in this manner 
again a divergence which
is hidden in Feynman parameters. This can be handled by splitting into $f(u)$
and $f(1)$ terms as above. For

\begin{equation}
\int \frac{d^nk}{(2 \pi)^n} \frac{ \mu^\epsilon \pi(k^2,m^2)}{(k_0+i \varepsilon)^2(k^2-\lambda^2
+i \varepsilon)
((k+q)^2-\lambda^2 +i \varepsilon)} \label{eq:clint}
\end{equation}

we find

\begin{eqnarray}
f(u) &=& 16 \; sin \left( \frac{\epsilon}{2}
\pi \right) 
\frac{ \Gamma \left( -1 - \frac{\epsilon}{2} \right) \Gamma \left(
1 + \epsilon \right) \Gamma \left( 2 + \frac{\epsilon}{2} \right) \eta^
\epsilon}{ (4 \pi)^4 \pi \; m^2} \int^1_0 dx \int^1_0 dv \frac{x (1-x) \;u}{ \alpha^{2+
\frac{\epsilon}{2}} \left( \Delta + \frac{\lambda^2}{m^2} u \right)^{1+
\epsilon}} \label{eq:clk0fu} \\
f(1) &=& 16 \;  sin \left( \frac{\epsilon}{2}
\pi \right) 
\frac{ \Gamma \left( -1 - \frac{\epsilon}{2} \right) \Gamma \left(
1 + \epsilon \right) \Gamma \left( 2 + \frac{\epsilon}{2} \right) \eta^
\epsilon}{ (4 \pi)^4 \pi \; m^2} \int^1_0 dx \int^1_0 dv \frac{x (1-x)}{ 
\left( \frac{-q^2}{m^2} v (1-v) + \frac{\lambda^2}{m^2} \right)^{1+
\epsilon}} \label{eq:clk0f1} 
\end{eqnarray}

The vertex correction graph ${\cal M}_{vc_3}$ and the integral occurring in
the onle-loop verex correction term ${\cal M}_{olvc}$ can
be calculated analogously. Here we have

\begin{equation}
\int \frac{d^nk}{(2 \pi)^n} \frac{ \mu^\epsilon \pi(k^2,m^2)}{(k_0+i \varepsilon)^2(k^2-\lambda^2
+i \varepsilon)
} \label{eq:vc3int}
\end{equation}

with the corresponding solutions

\begin{eqnarray}
f(u) &=& -16 \; sin \left( \frac{\epsilon}{2}
\pi \right) 
\frac{ \Gamma \left( - \frac{\epsilon}{2} \right) \Gamma \left(
\epsilon \right) \Gamma \left( 1 + \frac{\epsilon}{2} \right) \eta^
\epsilon}{ (4 \pi)^4 \pi \; q^2} \int^1_0 dx \frac{x (1-x)}{ \alpha^{1+
\frac{\epsilon}{2}} \left( 1-u + \frac{\lambda^2}{m^2} u \right)^{
\epsilon}} \label{eq:vc3k0fu} \\
f(1) &=& - \frac{8}{3} \; sin \left( \frac{\epsilon}{2}
\pi \right) 
\frac{ \Gamma \left( - \frac{\epsilon}{2} \right) \Gamma \left(
\epsilon \right) \Gamma \left( 1 + \frac{\epsilon}{2} \right) \eta^
\epsilon}{ (4 \pi)^4 \pi \; q^2}  
\left( \frac{m^2}{\lambda^2} \right)^{
\epsilon} \label{eq:vc3k0f1} 
\end{eqnarray}

\section*{Acknowledgments}
The author would like to thank S.J. Brodsky for
very valuable discussions in the early stages of the paper.
G. Buchalla  and M. Peter are greatfully acknowledged  
for helpful communications
on technical aspects of this work. Thanks also go to T. Hurth,
M. Wuesthoff, M. Heyssler
and J. Rathsman for
constructive conversations on several matters.

\bibliographystyle{plain}
\bibliography{books,hqp,Brodsky,beta,genthy,tli,melles}
\end{document}